\documentclass[twocolumn,letterpaper,aps,prd,superscriptaddress,showpacs,nofootinbib,floatfix]{revtex4-1}

\usepackage{graphicx}	
\usepackage{xspace}     
\usepackage{amsmath}

\newcommand{\pt}{\mbox{$p_{T}$}\xspace}
\newcommand{\all}{\mbox{$A_{LL}$}\xspace}

\newcommand{\pzall}{\mbox{$A_{LL}^{\pi^{0}}$}\xspace}
\newcommand{\etall}{\mbox{$A_{LL}^{\eta}$}\xspace}
\newcommand{\dg}{\mbox{$\Delta G$}\xspace}
\newcommand{\pz}{\mbox{$\pi^{0}$}\xspace}
\newcommand{\et}{\mbox{$\eta$}\xspace}
\newcommand{\punit}{GeV/$c$\xspace}
\newcommand{\munit}{MeV/$c^{2}$\xspace}
\newcommand{\gunit}{$\mbox{~GeV}^2$\xspace}
\newcommand{\mgg}{\mbox{$m_{\gamma\gamma}$}\xspace}

\newcommand{\sqs}{\mbox{$\sqrt{s}$}\xspace}

\begin{document}


\title{Inclusive double-helicity asymmetries in neutral-pion and eta-meson 
production in $\vec{p}+\vec{p}$ collisions at $\sqrt{s} = 200$~GeV}

\newcommand{\abilene}{Abilene Christian University, Abilene, Texas 79699, USA}
\newcommand{\augie}{Department of Physics, Augustana College, Sioux Falls, South Dakota 57197, USA}
\newcommand{\banaras}{Department of Physics, Banaras Hindu University, Varanasi 221005, India}
\newcommand{\barc}{Bhabha Atomic Research Centre, Bombay 400 085, India}
\newcommand{\baruch}{Baruch College, City University of New York, New York, New York, 10010 USA}
\newcommand{\bnlcoll}{Collider-Accelerator Department, Brookhaven National Laboratory, Upton, New York 11973-5000, USA}
\newcommand{\bnlphys}{Physics Department, Brookhaven National Laboratory, Upton, New York 11973-5000, USA}
\newcommand{\caucr}{University of California - Riverside, Riverside, California 92521, USA}
\newcommand{\charlesczech}{Charles University, Ovocn\'{y} trh 5, Praha 1, 116 36, Prague, Czech Republic}
\newcommand{\chonbuk}{Chonbuk National University, Jeonju, 561-756, Korea}
\newcommand{\ciae}{Science and Technology on Nuclear Data Laboratory, China Institute of Atomic Energy, Beijing 102413, P.~R.~China}
\newcommand{\cns}{Center for Nuclear Study, Graduate School of Science, University of Tokyo, 7-3-1 Hongo, Bunkyo, Tokyo 113-0033, Japan}
\newcommand{\colorado}{University of Colorado, Boulder, Colorado 80309, USA}
\newcommand{\columbia}{Columbia University, New York, New York 10027 and Nevis Laboratories, Irvington, New York 10533, USA}
\newcommand{\czechtech}{Czech Technical University, Zikova 4, 166 36 Prague 6, Czech Republic}
\newcommand{\dapnia}{Dapnia, CEA Saclay, F-91191, Gif-sur-Yvette, France}
\newcommand{\elte}{ELTE, E{\"o}tv{\"o}s Lor{\'a}nd University, H - 1117 Budapest, P{\'a}zm{\'a}ny P. s. 1/A, Hungary}
\newcommand{\ewha}{Ewha Womans University, Seoul 120-750, Korea}
\newcommand{\fsu}{Florida State University, Tallahassee, Florida 32306, USA}
\newcommand{\gsu}{Georgia State University, Atlanta, Georgia 30303, USA}
\newcommand{\hanyang}{Hanyang University, Seoul 133-792, Korea}
\newcommand{\hiroshima}{Hiroshima University, Kagamiyama, Higashi-Hiroshima 739-8526, Japan}
\newcommand{\ihepprot}{IHEP Protvino, State Research Center of Russian Federation, Institute for High Energy Physics, Protvino, 142281, Russia}
\newcommand{\illuiuc}{University of Illinois at Urbana-Champaign, Urbana, Illinois 61801, USA}
\newcommand{\inrras}{Institute for Nuclear Research of the Russian Academy of Sciences, prospekt 60-letiya Oktyabrya 7a, Moscow 117312, Russia}
\newcommand{\instpasczech}{Institute of Physics, Academy of Sciences of the Czech Republic, Na Slovance 2, 182 21 Prague 8, Czech Republic}
\newcommand{\isu}{Iowa State University, Ames, Iowa 50011, USA}
\newcommand{\jaea}{Advanced Science Research Center, Japan Atomic Energy Agency, 2-4 Shirakata Shirane, Tokai-mura, Naka-gun, Ibaraki-ken 319-1195, Japan}
\newcommand{\jyvaskyla}{Helsinki Institute of Physics and University of Jyv{\"a}skyl{\"a}, P.O.Box 35, FI-40014 Jyv{\"a}skyl{\"a}, Finland}
\newcommand{\kek}{KEK, High Energy Accelerator Research Organization, Tsukuba, Ibaraki 305-0801, Japan}
\newcommand{\korea}{Korea University, Seoul, 136-701, Korea}
\newcommand{\kurchatov}{Russian Research Center ``Kurchatov Institute", Moscow, 123098 Russia}
\newcommand{\kyoto}{Kyoto University, Kyoto 606-8502, Japan}
\newcommand{\labllr}{Laboratoire Leprince-Ringuet, Ecole Polytechnique, CNRS-IN2P3, Route de Saclay, F-91128, Palaiseau, France}
\newcommand{\lahorelums}{Physics Department, Lahore University of Management Sciences, Lahore, Pakistan}
\newcommand{\lawllnl}{Lawrence Livermore National Laboratory, Livermore, California 94550, USA}
\newcommand{\losalamos}{Los Alamos National Laboratory, Los Alamos, New Mexico 87545, USA}
\newcommand{\lpc}{LPC, Universit{\'e} Blaise Pascal, CNRS-IN2P3, Clermont-Fd, 63177 Aubiere Cedex, France}
\newcommand{\lund}{Department of Physics, Lund University, Box 118, SE-221 00 Lund, Sweden}
\newcommand{\maryland}{University of Maryland, College Park, Maryland 20742, USA}
\newcommand{\mass}{Department of Physics, University of Massachusetts, Amherst, Massachusetts 01003-9337, USA }
\newcommand{\michigan}{Department of Physics, University of Michigan, Ann Arbor, Michigan 48109-1040, USA}
\newcommand{\muenster}{Institut fur Kernphysik, University of Muenster, D-48149 Muenster, Germany}
\newcommand{\muhlenberg}{Muhlenberg College, Allentown, Pennsylvania 18104-5586, USA}
\newcommand{\myongji}{Myongji University, Yongin, Kyonggido 449-728, Korea}
\newcommand{\nagasaki}{Nagasaki Institute of Applied Science, Nagasaki-shi, Nagasaki 851-0193, Japan}
\newcommand{\newmex}{University of New Mexico, Albuquerque, New Mexico 87131, USA }
\newcommand{\nmsu}{New Mexico State University, Las Cruces, New Mexico 88003, USA}
\newcommand{\ohio}{Department of Physics and Astronomy, Ohio University, Athens, Ohio 45701, USA}
\newcommand{\ornl}{Oak Ridge National Laboratory, Oak Ridge, Tennessee 37831, USA}
\newcommand{\orsay}{IPN-Orsay, Universite Paris Sud, CNRS-IN2P3, BP1, F-91406, Orsay, France}
\newcommand{\peking}{Peking University, Beijing 100871, P.~R.~China}
\newcommand{\pnpi}{PNPI, Petersburg Nuclear Physics Institute, Gatchina, Leningrad region, 188300, Russia}
\newcommand{\riken}{RIKEN Nishina Center for Accelerator-Based Science, Wako, Saitama 351-0198, Japan}
\newcommand{\rikjrbrc}{RIKEN BNL Research Center, Brookhaven National Laboratory, Upton, New York 11973-5000, USA}
\newcommand{\rikkyo}{Physics Department, Rikkyo University, 3-34-1 Nishi-Ikebukuro, Toshima, Tokyo 171-8501, Japan}
\newcommand{\saopaulo}{Universidade de S{\~a}o Paulo, Instituto de F\'{\i}sica, Caixa Postal 66318, S{\~a}o Paulo CEP05315-970, Brazil}
\newcommand{\seoulnat}{Seoul National University, Seoul, Korea}
\newcommand{\stonybrkc}{Chemistry Department, Stony Brook University, SUNY, Stony Brook, New York 11794-3400, USA}
\newcommand{\stonycrkp}{Department of Physics and Astronomy, Stony Brook University, SUNY, Stony Brook, New York 11794-3800, USA}
\newcommand{\tenn}{University of Tennessee, Knoxville, Tennessee 37996, USA}
\newcommand{\titech}{Department of Physics, Tokyo Institute of Technology, Oh-okayama, Meguro, Tokyo 152-8551, Japan}
\newcommand{\tsukuba}{Institute of Physics, University of Tsukuba, Tsukuba, Ibaraki 305, Japan}
\newcommand{\vandy}{Vanderbilt University, Nashville, Tennessee 37235, USA}
\newcommand{\weizmann}{Weizmann Institute, Rehovot 76100, Israel}
\newcommand{\wigner}{Institute for Particle and Nuclear Physics, Wigner Research Centre for Physics, Hungarian Academy of Sciences (Wigner RCP, RMKI) H-1525 Budapest 114, POBox 49, Budapest, Hungary}
\newcommand{\yonsei}{Yonsei University, IPAP, Seoul 120-749, Korea}
\affiliation{\abilene}
\affiliation{\augie}
\affiliation{\banaras}
\affiliation{\barc}
\affiliation{\baruch}
\affiliation{\bnlcoll}
\affiliation{\bnlphys}
\affiliation{\caucr}
\affiliation{\charlesczech}
\affiliation{\chonbuk}
\affiliation{\ciae}
\affiliation{\cns}
\affiliation{\colorado}
\affiliation{\columbia}
\affiliation{\czechtech}
\affiliation{\dapnia}
\affiliation{\elte}
\affiliation{\ewha}
\affiliation{\fsu}
\affiliation{\gsu}
\affiliation{\hanyang}
\affiliation{\hiroshima}
\affiliation{\ihepprot}
\affiliation{\illuiuc}
\affiliation{\inrras}
\affiliation{\instpasczech}
\affiliation{\isu}
\affiliation{\jaea}
\affiliation{\jyvaskyla}
\affiliation{\kek}
\affiliation{\korea}
\affiliation{\kurchatov}
\affiliation{\kyoto}
\affiliation{\labllr}
\affiliation{\lahorelums}
\affiliation{\lawllnl}
\affiliation{\losalamos}
\affiliation{\lpc}
\affiliation{\lund}
\affiliation{\maryland}
\affiliation{\mass}
\affiliation{\michigan}
\affiliation{\muenster}
\affiliation{\muhlenberg}
\affiliation{\myongji}
\affiliation{\nagasaki}
\affiliation{\newmex}
\affiliation{\nmsu}
\affiliation{\ohio}
\affiliation{\ornl}
\affiliation{\orsay}
\affiliation{\peking}
\affiliation{\pnpi}
\affiliation{\riken}
\affiliation{\rikjrbrc}
\affiliation{\rikkyo}
\affiliation{\saopaulo}
\affiliation{\seoulnat}
\affiliation{\stonybrkc}
\affiliation{\stonycrkp}
\affiliation{\tenn}
\affiliation{\titech}
\affiliation{\tsukuba}
\affiliation{\vandy}
\affiliation{\weizmann}
\affiliation{\wigner}
\affiliation{\yonsei}
\author{A.~Adare} \affiliation{\colorado}
\author{C.~Aidala} \affiliation{\losalamos} \affiliation{\michigan}
\author{N.N.~Ajitanand} \affiliation{\stonybrkc}
\author{Y.~Akiba} \affiliation{\riken} \affiliation{\rikjrbrc}
\author{R.~Akimoto} \affiliation{\cns}
\author{H.~Al-Ta'ani} \affiliation{\nmsu}
\author{J.~Alexander} \affiliation{\stonybrkc}
\author{K.R.~Andrews} \affiliation{\abilene}
\author{A.~Angerami} \affiliation{\columbia}
\author{K.~Aoki} \affiliation{\riken}
% ...
\collaboration{PHENIX Collaboration} \noaffiliation

\date{\today}


\begin{abstract}


Results are presented from data recorded in 2009 by the PHENIX experiment
at the Relativistic Heavy Ion Collider for the double-longitudinal spin
asymmetry, $A_{LL}$, for $\pi^0$ and $\eta$ production in $\sqrt{s} =
200$~GeV polarized $p$$+$$p$ collisions. Comparison of the $\pi^0$ results
with different theory expectations based on fits of other published 
data showed a preference for small positive values of gluon
polarization, $\Delta G$, in the proton in the probed Bjorken $x$ range.
The effect of adding the new 2009 \pz data to a recent global analysis of
polarized scattering data is also shown, resulting in a best fit value
$\Delta G^{[0.05,0.2]}_{\mbox{DSSV}} = 0.06^{+0.11}_{-0.15}$ in the range
$0.05<x<0.2$, with the uncertainty at $\Delta \chi^2 = 9$ when considering
only statistical experimental uncertainties.  Shifting the PHENIX data
points by their systematic uncertainty leads to a variation of the
best-fit value of $\Delta G^{[0.05,0.2]}_{\mbox{DSSV}}$ between $0.02$ and
$0.12$, demonstrating the need for full treatment of the experimental
systematic uncertainties in future global analyses.

\end{abstract}


\maketitle

\section{Introduction}

The proton has a finite charge radius and can be described as a collection 
of fermionic quarks whose interaction is mediated by bosonic gluons.  The 
proton is also a spin-1/2 fermion itself, which constrains the total 
angular momentum of these constituents and have been described in several 
proposed sum rules~\cite{ref:sum_rules1,ref:sum_rules_chen, 
ref:sum_rules_hatta, ref:sum_rules_ji, ref:sum_rules_wakamatsu}.  In the 
infinite momentum frame, all possible contributions to the proton spin can 
be classified according to the Manohar-Jaffe sum 
rule~\cite{ref:sum_rules1},

\begin{equation}
S_p = \frac{1}{2} = \frac{1}{2} \Delta \Sigma + \dg + L_q + L_g,
\label{e:sum_rule}
\end{equation}
which makes explicit the contributions from quark and gluon spin 
($\Delta\Sigma$ and $\dg$, respectively) and orbital angular momentum 
($L_q$ and $L_g$, respectively).

Early experiments discovered that the $\frac{1}{2} \Delta \Sigma$ term 
falls far short of the total \cite{ref:EMC:spincrisis2, ref:NA47, 
ref:E143}.  Current knowledge from global fits~\cite{ref:DSSV1, ref:DSSV2, 
ref:LSS, ref:BB10, ref:AAC} of polarized deeply-inelastic-scattering (DIS) 
and semi-inclusive DIS (SIDIS) data~\cite{ref:NA47, ref:E143, 
ref:strangeSIDIS_COMPASS, ref:strangeSIDIS_HERMES} puts the contribution 
at only 25-35\% of the proton spin, depending on the assumptions used, 
including whether SU(3) symmetry is enforced.  This realization provided 
the motivation to study the $\dg$ term by colliding longitudinally 
polarized protons at the Relativistic Heavy Ion Collider (RHIC), including 
the results presented here.

\begin{figure}[htb]
\includegraphics[width=1.0\linewidth]{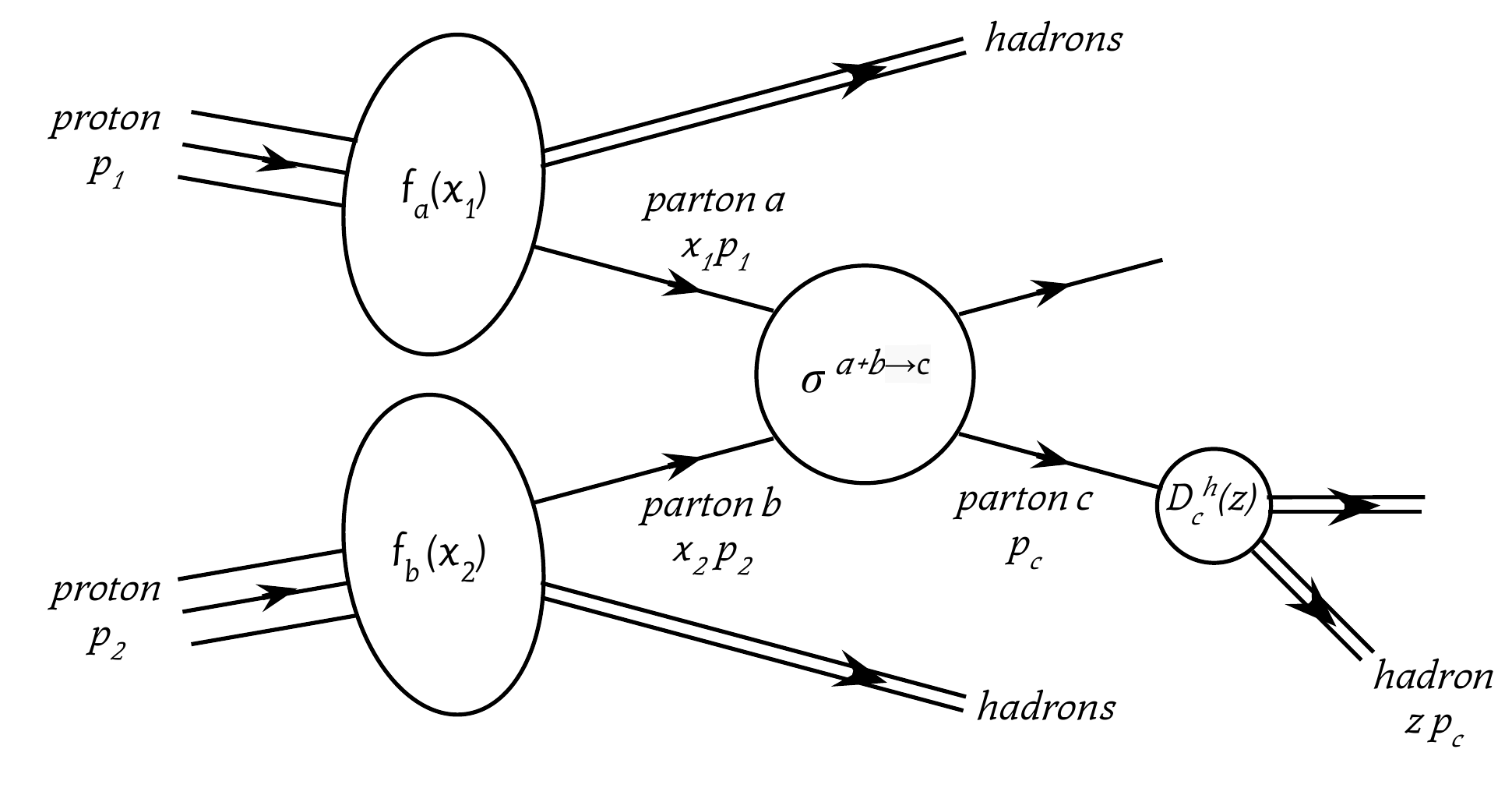}
\caption{
Diagram showing the three elements of pQCD factorization: parton 
distribution functions $f_{a,b}(x)$, partonic cross sections 
$\sigma^{a+b \rightarrow c}$, and fragmentation functions $D_c^h(z)$.}
\label{f:p_p_feynman}
\end{figure}

Polarized proton collisions at RHIC allow access to \dg at leading order 
(LO) in perturbative quantum chromodynamics (pQCD), unlike lepton-hadron 
scattering experiments that are only sensitive to \dg via photon-gluon 
fusion at next-to-leading order (NLO) in pQCD or via 
momentum-transfer-scaling violations of the polarized structure function 
$g_1$.  RHIC experiments make the connection to \dg via inclusive 
double-helicity asymmetries, \all, defined by
\begin{equation}
  \all = \frac{\Delta\sigma}{\sigma} = \frac{\sigma_{++} - \sigma_{+-}}{\sigma_{++} + \sigma_{+-}}.
  \label{eq:allsigma}
\end{equation}
Here, $\sigma$ ($\Delta \sigma$) is the (polarized) cross section for a 
given observable, and `$++$' (`$+-$') signifies $\vec{p}+\vec{p}$ 
collisions with the same (opposite) helicity.  Within the framework of 
pQCD, \all can also be ``factorized'' to make the parton spin 
contributions explicit:
\begin{widetext}
  \begin{equation}
    \all = \frac{\sum_{abc}\Delta f_a(x_1,\mu_F^2) \otimes \Delta f_b(x_2,\mu_F^2) \otimes \Delta\sigma^{a+b\rightarrow c+X}(x_1,x_2,p_c,\mu_F^2,\mu_R^2,\mu_{FF}^2) \otimes D^{h}_{c}(z,\mu_{FF}^2)}{\sum_{abc} f_a(x_1,\mu_F^2) \otimes f_b(x_2,\mu_F^2) \otimes \sigma^{a+b\rightarrow c+X}(x_1,x_2,p_c,\mu_F^2,\mu_R^2,\mu_{FF}^2) \otimes D^{h}_{c}(z,\mu_{FF}^2)},
    \label{eq:alltheory}
  \end{equation}
\end{widetext} 
where $f_{a,b}$ ($\Delta f_{a,b}$) are the unpolarized (polarized) parton 
distribution functions [PDF (pPDF)], phenomenological functions describing the 
statistical distribution for partons $a,b$ (gluons, quarks, or antiquarks) 
in a proton as a function of the momentum fraction Bjorken $x$.  $D^h_c$ 
is the fragmentation function (FF) describing the probability for a parton 
$c$ with momentum $p_c$ to fragment into a hadron $h$ with momentum $p_h$ 
and thus with a given $z=p_h/p_c$.  $\Delta\sigma^{a+b\rightarrow c+X}$ 
and $\sigma^{a+b\rightarrow c+X}$ are the polarized and unpolarized 
partonic cross sections, respectively, and are calculable in pQCD.  
Factorization, renormalization and fragmentation scales $\mu_F$, $\mu_R$ 
and $\mu_{FF}$ are used to separate the perturbative and nonperturbative 
parts.  The diagram in Fig.~\ref{f:p_p_feynman} summarizes the components 
of pQCD factorization.  The theoretical calculations discussed in this 
work with respect to our results are performed at NLO in pQCD.

To test the applicability of NLO pQCD to our \all results, PHENIX 
has previously published \pz- and \et-meson cross sections 
\cite{ref:run5pi0xsect_all,ref:run6etaxsect_all}.  These cross sections, 
along with others at $\sqs=200$~GeV for jets 
\cite{ref:run6starjetxsect_all} and direct photons 
\cite{ref:run6directgammaxsect}, are well described by the theory within 
its uncertainties, based on the method of varying the choice of scales by 
a factor of two.  In our previous publication~\cite{ref:run6pi0_all}, we 
examined the impact of this theoretical scale uncertainty with respect to 
our $A_{LL}^{\pi^0}$ results, and found that it is important and should be 
considered in future global analyses.
 
A number of different channels can be used to access the gluon 
polarization using Eq.~\ref{eq:alltheory}, including a final state hadron 
or jet, as well as rarer probes such as direct photon and heavy flavor 
\cite{ref:run9elect_all}.  The latter of these are produced through fewer 
processes, which allows for a simple leading-order interpretation of the 
results.  Jets or low-mass hadrons such as pions are not as readily 
interpretable due to the multiple QCD processes through which they are 
produced, but they have significantly higher production rates.  PHENIX 
results for \pzall \cite{ref:run5pi0xsect_all, ref:run6pi0_all, 
ref:vern} and results for jet \all from the STAR Experiment at RHIC 
\cite{ref:run6starjetxsect_all, ref:run5starjetall} have ruled out large 
values of \dg but are still consistent with a range of assumptions, 
including fixing the polarized PDF for the gluon, $\Delta g(x,\mu^2)$, to 
zero at an NLO input scale of $\mu^2 = 0.40 \mbox{~GeV}^2$.  The 
constraining power of these results has been quantified via inclusion in a 
global fit of polarized DIS and semi-inclusive DIS results by de Florian, 
et. al (DSSV) \cite{ref:DSSV1, ref:DSSV2}, resulting in an integral 
$\Delta G^{[0.05,0.2]}_{\mbox{\sc dssv08}} = 0.005^{+0.129}_{-0.164}$ in the 
Bjorken-$x$ range $[0.05,0.2]$.  As detailed in~\cite{ref:run6pi0_all}, 
the full $x$ range probed by the PHENIX $A_{LL}^{\pi^0}$ measurements is 
$[0.02,0.3]$.

The $\etall$ has also been measured~\cite{ref:run6etaxsect_all}, but it 
has not yet been used in global fits.  One reason for this is that 
existing $e^+ + e^-$ data does not constrain $\eta$ fragmentation 
functions as well as those for the pions~\cite{ref:DSS_FF, ref:eta_FF}.  
However, PHENIX has released results for the $\eta / \pi^0$ cross section 
ratio in $p$$+$$p$ collisions \cite{ref:run6etaxsect_all, 
ref:eta_pi0_xsect_ratios} with systematic uncertainties much smaller than 
on either cross section measurement alone.  Future inclusion of this ratio 
in global fits could be used to circumvent this issue with the 
fragmentation functions.

In this paper, we present measurements of \all in \pz- and 
\et-meson production 
in longitudinally polarized $p$$+$$p$ collisions at $\sqs=200$~GeV, based on 
data collected in 2009, which approximately doubles the 
statistics in the $\sqs=200$~GeV PHENIX neutral meson \all dataset 
~\cite{ref:run6pi0_all, ref:run6etaxsect_all} and extends the measured 
$p_T$ range.  Descriptions of RHIC and the PHENIX experiment are laid out 
in Section~\ref{s:experiment}, followed by a detailed account of the 
analysis procedure in Section~\ref{s:Analysis} including discussion of the 
systematic uncertainties.  Finally, in Sections~\ref{s:results} 
and~\ref{s:discussion}, we show our final results and discuss them in 
relation to global analyses of polarized scattering data.

\section{Experimental Setup}
\label{s:experiment}

\subsection{Polarized Beams at RHIC}
\label{s:rhic}

RHIC comprises two counter-rotating storage rings, designated blue and 
yellow, in each of which as many as 120 polarized proton bunches of 
$10^{11}$ protons or more can be accelerated to an energy of $255$~GeV per 
proton.  In the 2009 run, RHIC was typically operated with 109 filled 
bunches in each ring.  The rings intersect in 6 locations such that the 
bunches collide with a one-to-one correspondence.  This allows an 
unambiguous definition of 120 ``crossings'' per revolution at each 
experiment, with a 106~ns separation.  At PHENIX, there were 107 crossings 
in which both bunches were filled and 4 crossings with only the bunch in 
one ring filled to enable study of beam background.

Outside of the experimental interaction regions, the stable polarization 
direction in RHIC is vertical~\cite{ref:rhicnim}.  The polarization for 
each bunch can be aligned or anti-aligned with this vertical axis at 
injection, allowing for variation over all four possible polarization 
combinations within 4 crossings, or 424~ns.  To cancel false asymmetries 
related to coupling between the polarization patterns and the 
bunch/crossing structure, four different polarization vs. bunch patterns, 
hereafter referred to as ``spin patterns,'' were used, defined by changing 
the sign of all polarizations in one or both beams from the base pattern.  
The patterns were cycled after each successful beam store, or ``fill''.

Determination of the beam polarizations required combining measurements 
from two separate polarimeters.  First, the relative beam polarizations 
were measured several times per fill using a fast, high-statistics 
relative polarimeter, which detects elastic scattering off of a thin 
carbon target that is moved across the beam.  This polarimeter can 
determine both the relative magnitude of the polarization and any 
variation across the width of the beam \cite{ref:cni}.  This measurement 
was normalized by comparing its average over the entire dataset to the 
average of an absolute polarization measurement from the second 
polarimeter, which is based on scattering of the beam with a 
continuously-running polarized hydrogen gas-jet target \cite{ref:hjet}.  
For 2009 $\sqrt{s} = 200$~GeV running, the average beam polarizations were 
$P_B = 0.56$ for the blue beam and $P_Y = 0.55$ for the yellow beam, for a 
product $P_B P_Y = 0.31$.  The overall relative scale uncertainty on the 
product $P_B P_Y$ was $6.5\%$, with $4.8\%$ of that considered correlated 
with the polarization uncertainties from RHIC runs in 2005 and 
2006~\cite{ref:CADnote}.

\subsection{The PHENIX Experiment}
\label{s:phenix}

\begin{figure}
\includegraphics[width=1.0\linewidth]{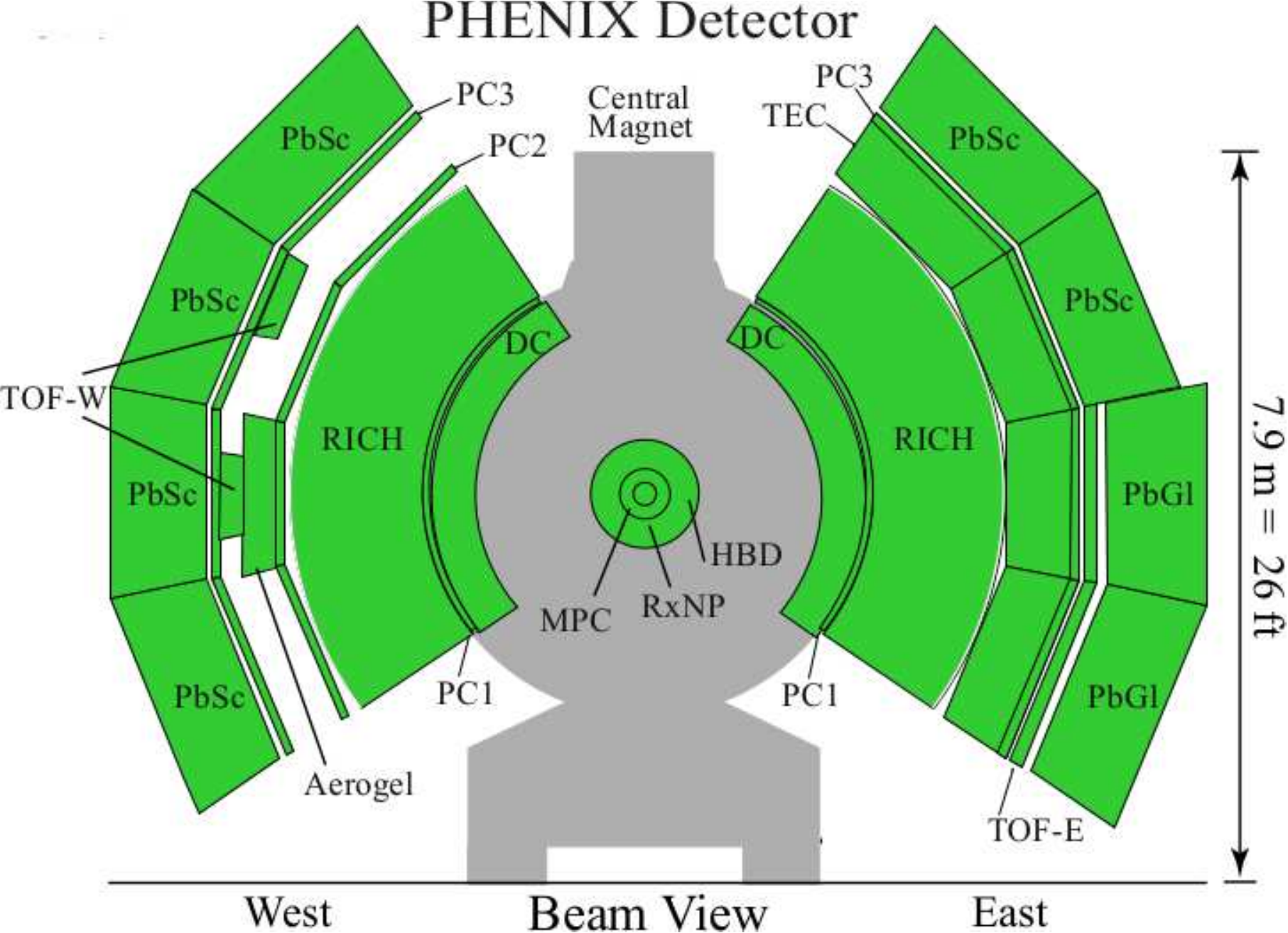} 
\caption{(color online)
Cross section of the PHENIX Central Arms in their 2009-2010 configuration.  
The EMCal (PbSc and PbGl) is the primary subsystem used in this analysis.  
The PC3 is also used to veto charged particles.  Not pictured in this view 
are the BBC and ZDC luminosity monitors at forward rapidity.}
\label{f:Phenix_2009}
\end{figure}

The PHENIX detector \cite{ref:phenixnim} comprises two forward muon arms 
and two central arms, shown in Fig.~\ref{f:Phenix_2009}.  Except for 
luminosity normalization using counters at forward rapidity, the present 
analysis uses only the central arms, each of which cover a pseudorapidity 
range of $|\et|<0.35$ and have azimuthal coverage of 
$\Delta\phi=\frac{\pi}{2}$.  The PHENIX central magnet comprises two coils 
which provide a field-integral of up to $1.15$~Tm in $|\et|<0.35$ when 
they are run with the same polarity, as was done in 2005 and 2006. In 
2009, the two central coils were run with opposite polarity to create a 
field free region near the beam pipe for the newly installed hadron-blind 
detector \cite{ref:hbdnim}, which is not used in the present analysis and 
has a negligible effect on \pz- and \et-meson decays as a conversion 
material.  From a radius of 2--5~m, which is outside the magnetic field 
region, there are several tracking and particle-identification detectors 
that are not used in this analysis.  At a radius of approximately 5~m, 
there is a thin multiwire proportional chamber called the pad chamber 
(PC3) followed immediately by an electromagnetic calorimeter (EMCal), both 
of which are used in this analysis.

\subsubsection{EMCal}

Neutral pion and eta mesons can both be analyzed via their diphoton decay 
channel (for the $\pi^0$, the branching ratio is 99\%, for the $\eta$, 
39\% \cite{ref:PDG}), which allows for accurate reconstruction of both 
mesons using a sufficiently segmented electromagnetic calorimeter.
The PHENIX EMCal employs two separate technologies to have 
sensitivity to calorimeter-based systematic effects.  Six out of the eight 
EMCal sectors are lead scintillator (PbSc), which are Shashlik 
calorimeters based on scintillation calorimetry, while the remaining two 
are lead-glass (PbGl), which are based on \v{C}erenkov radiation 
calorimetry, which makes them significantly less responsive to hadrons.

Both the PbSc and PbGl are designed to measure the total energy of an 
electromagnetic shower, with active depths of $18.8$ and $14.3$ radiation 
lengths, respectively.  The nominal energy resolutions from test-beam data 
are $8.1\%/\sqrt{E[\mbox{GeV}]} \oplus 2.1\%$ and 
$6.0\%/\sqrt{E[\mbox{GeV}]} \oplus 0.9\%$~\cite{ref:emcalnim}.

The PbSc (PbGl) also have sufficient lateral tower segmentation, 
$\Delta~\eta~\sim$~0.01~(0.008) and $\Delta~\phi~\sim$~0.01~(0.008)~rad, 
to measure not only the position, but also the transverse distribution of 
an electromagnetic shower, with a typical shower contained in a 3$\times$3 
array of EMCal towers.  The segmentation is also sufficient to avoid 
pile-up at the highest RHIC $p$$+$$p$ rates and in the high-multiplicity 
environment of heavy ion collisions.

The relative time-of-flight (ToF) for showers can also be measured with 
the EMCal with a timing resolution of about $0.7$~ns for the present data.  
This measurement can be used to reduce the contribution from hadrons and 
other backgrounds that are out of time from the expected arrival for a 
photon.

\subsubsection{EMCal Trigger}

To record a significant sample of events containing a \pz or \et meson 
with large transverse momentum ($p_T$), a high energy photon trigger is 
used.  A trigger tile is defined as a $2\times2$ array of EMCal towers, 
and, for the present analysis, the energy in a $2\times2$ array of tiles 
(or $4\times4$ towers) is summed and compared to the trigger threshold.  
To reduce loss at the edge of a tile, these groups of $4\times4$ towers 
overlap.  For this analysis, we use two trigger thresholds, one at 1.4~GeV 
and one at 2.1~GeV.  For diphoton decays, these are maximally efficient 
at parent meson energies of $>4$~GeV and $>6$~GeV, respectively.  Since 
the reset time of the trigger, $\sim$ 140~ns, is longer than the $\sim$ 
106~ns between bunches, two separate trigger circuits are used to read out 
even and odd-numbered crossings.  This can lead to variations in the 
effective thresholds in even and odd crossings, requiring the analysis to 
be done separately for each.

\subsubsection{PC3}

The PC3 provides nonprojective tracking via a pixelated cathode that is 
segmented into $16.8$~mm~$\times$~$16.8$~mm pads, giving it excellent 
spatial resolution.  This detector is used in the present analyses only as 
a veto for charged particle clusters, as described in 
Section~\ref{ss:ch_vet}.

\subsubsection{Luminosity Monitors}

PHENIX has two luminosity monitors with which to normalize the luminosity 
variations between same and opposite helicity bunch crossings.  The main 
luminosity monitor is the beam-beam counter (BBC)~\cite{ref:phenixnim}, 
which comprises two arms located $|z|=144$~cm from the interaction-point 
(IP) along the beam axis, covering a pseudorapidity range of 
$3.1<|\eta|<3.9$.  Each arm has 64 quartz crystal \v{C}erenkov radiators 
attached to photomultiplier tubes.  The BBC also functions as the 
minimum-bias (MB) collision trigger for this dataset, with a requirement 
that at least one photomultiplier tube fire in each arm and that the 
timing of the hits reconstructs to a collision with a $z$-vertex within 
30~cm of the nominal IP.  The yield of MB triggers in crossings where the 
data acquisition system was ready to take data was used to determine the 
luminosity.

The second luminosity monitor, the zero-degree calorimeter 
(ZDC)~\cite{ref:phenixnim}, comprises two arms located $|z|=18$~m from 
the IP along the beam axis, covering $|\eta|>6$.  Each arm is composed of 
three sections of hadron calorimeter composed of optical fibers for 
\v{C}erenkov sampling sandwiched between layers of tungsten absorber.  
The three sections constitute a total of 5 nuclear interaction lengths.  
As the arms lie beyond the bending magnets, which serve to separate the 
two beams outside the experimental area but also sweep away charged 
particles from the interaction, the ZDC primarily triggers on neutrals.  
A ZDC trigger requires a minimum energy deposit in each arm of nominally 
20~GeV.

\subsubsection{Local Polarimeter}

The \all measurements require longitudinal polarization.  Four spin 
rotator magnets (two in each ring) located outside of the PHENIX 
interaction region rotate the beam polarization from the stable vertical 
direction to the longitudinal direction before the IP and back to vertical 
afterward.  A position-sensitive shower-maximum detector, composed of 
vertical and horizontal scintillator strips, is located between the first 
and second sections in each ZDC arm.  It is used in conjunction with the 
ZDC to measure an azimuthal asymmetry in forward neutron production with a 
magnitude of 0.07~\cite{ref:neutronasymm} during transverse polarization 
running (with the spin rotators turned off).  This asymmetry should vanish 
when the beam polarization vector is perfectly longitudinal.  The size of 
the residual asymmetry can therefore be used to determine the remaining 
transverse component, and thus the degree of effective longitudinal 
polarization.  In 2009 at $\sqrt{s} = 200$~GeV, the fraction of the 
polarization along the longitudinal direction in the blue beam was 
$0.994^{+0.006}_{-0.008}\mbox{(stat)} ^{+0.003}_{-0.010}\mbox{(syst)}$ and 
in the yellow beam 
$0.974^{+0.014}_{-0.018}\mbox{(stat)}^{+0.019}_{-0.035}\mbox{(syst)}$.

\section{Data Analysis}
\label{s:Analysis}

\subsection{Event and Photon Selection}
\label{ss:cuts}

Events used in this analysis require a MB trigger in coincidence with a 
high energy trigger in the EMCal.  An offline vertex cut is applied, which 
requires that the vertex reconstructed using the BBC be within $|z|=30$~cm 
of the nominal IP.  On the order of two billion events passing this 
offline cut were analyzed.

Photon candidates are selected from all energy deposit clusters in the 
EMCal.  A minimum energy of 100~MeV in PbSc and 200~MeV in PbGl is 
required to reduce the impact of noise in the detector.  
Clusters centered on towers that are markedly noisy or dead, or centered 
on towers neighboring a markedly noisy or dead tower, are discarded.  
Clusters centered within two towers of the edge of each EMCal sector's 
acceptance are also excluded.

A major source of background in the photon candidate sample are charged 
hadrons, which are removed by three cuts based on shower shape, time of 
flight (ToF) and association with hits in the PC3.  For the shower shape 
cut, the distribution among towers of the total energy deposited is 
compared with the expected distribution for an electromagnetic shower, 
based on results from test beam data.  The resulting cut is 98\% efficient 
for photons.  The other two cuts are discussed in more detail below.

Also of concern is background of clusters from previous events; since they 
can be from crossings with a different bunch helicity combination, the 
asymmetries are affected.  Photons from meson decays in previous events 
are effectively removed by the trigger requirement described in 
Section~\ref{ss:meson_sel}.  The ToF cut is effective in targeting the 
remaining clusters of this type.

\subsubsection{Charge Veto Cut}
\label{ss:ch_vet}

One method to remove charged hadrons is to veto photon candidates with 
associated (charged particle) hits in the PC3.  However, to not  
unnecessarily remove real photons that pair-converted before the EMCal, 
but outside of the magnetic field, a two-sided cut was developed.

\begin{figure}
\includegraphics[width=1.0\linewidth]{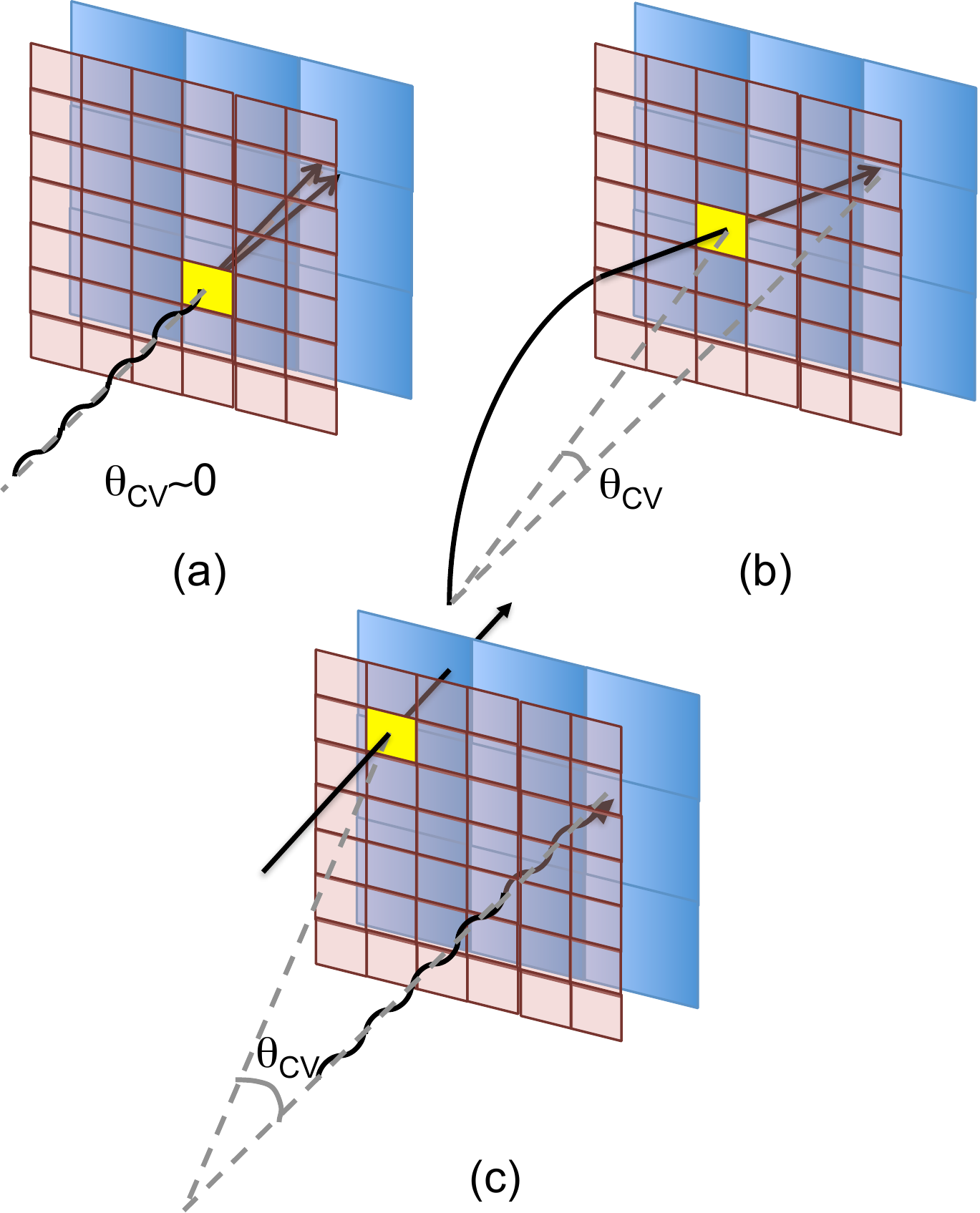}
\caption{(color online)
Schematic (not to scale) of the hits in the PC3 (transparent grid) and the 
related $\theta_{CV}$ from three particle classes, which leave clusters in 
the EMCal (solid grid behind PC3):  (a) photons that convert outside of 
the magnetic field prior to the EMCal, and have very small $\theta_{CV}$, 
(b) charged hadrons that bend in the magnetic field, and so have moderate 
sized $\theta_{CV}$, and (c) photons that do not convert, and are 
randomly associated with a different particle's PC3 hit, and therefore are 
likely to have large $\theta_{CV}$.}
\label{f:chargevetocartoon}
\end{figure}

We define two vectors: (1) the vector starting at the event vertex and 
pointing to a cluster in the EMCal and (2) the vector pointing from the 
vertex to the hit in the PC3 nearest to the EMCal cluster.  The angle 
between these vectors is defined as $\theta_{CV}$, the charge veto angle.  
The diagram in Fig.~\ref{f:chargevetocartoon} shows schematically how this 
angle is defined for three distinct cases, which can be classified 
according to the relative magnitude of $\theta_{CV}$:

\begin{enumerate}

\item Small $\theta_{CV}$: e$^+$e$^-$ pairs from photon conversions 
outside of the magnetic field region can still form a single cluster if 
their opening angle or the conversion's distance from the EMCal is small.  
In this case we may find an associated PC3 hit directly in front of the 
cluster, but we can still reconstruct the original photon from the energy 
deposited.  Thus we should retain clusters with small $\theta_{CV}$.

\item Moderate $\theta_{CV}$: Due to the separation between the PC3 and 
EMCal as well as the large EMCal penetration depth for hadrons compared to 
photons, it is not possible to draw a straight line connecting the EMCal 
cluster center, PC3 hit and collision vertex for charged hadrons that 
travel through (and bend in) the magnetic field.  Thus there will be some 
energy-dependent $\theta_{CV}$ region associated with these particles 
which can be used to exclude them from the analysis.

\item Large $\theta_{CV}$: The phase space for combinatorial association 
of an EMCal cluster with an unrelated PC3 hit increases linearly with 
$\tan(\theta_{CV})$.  Thus random association dominates this region and we 
should not throw out these clusters.

\end{enumerate}

\begin{figure}
\includegraphics[width=1.0\linewidth]{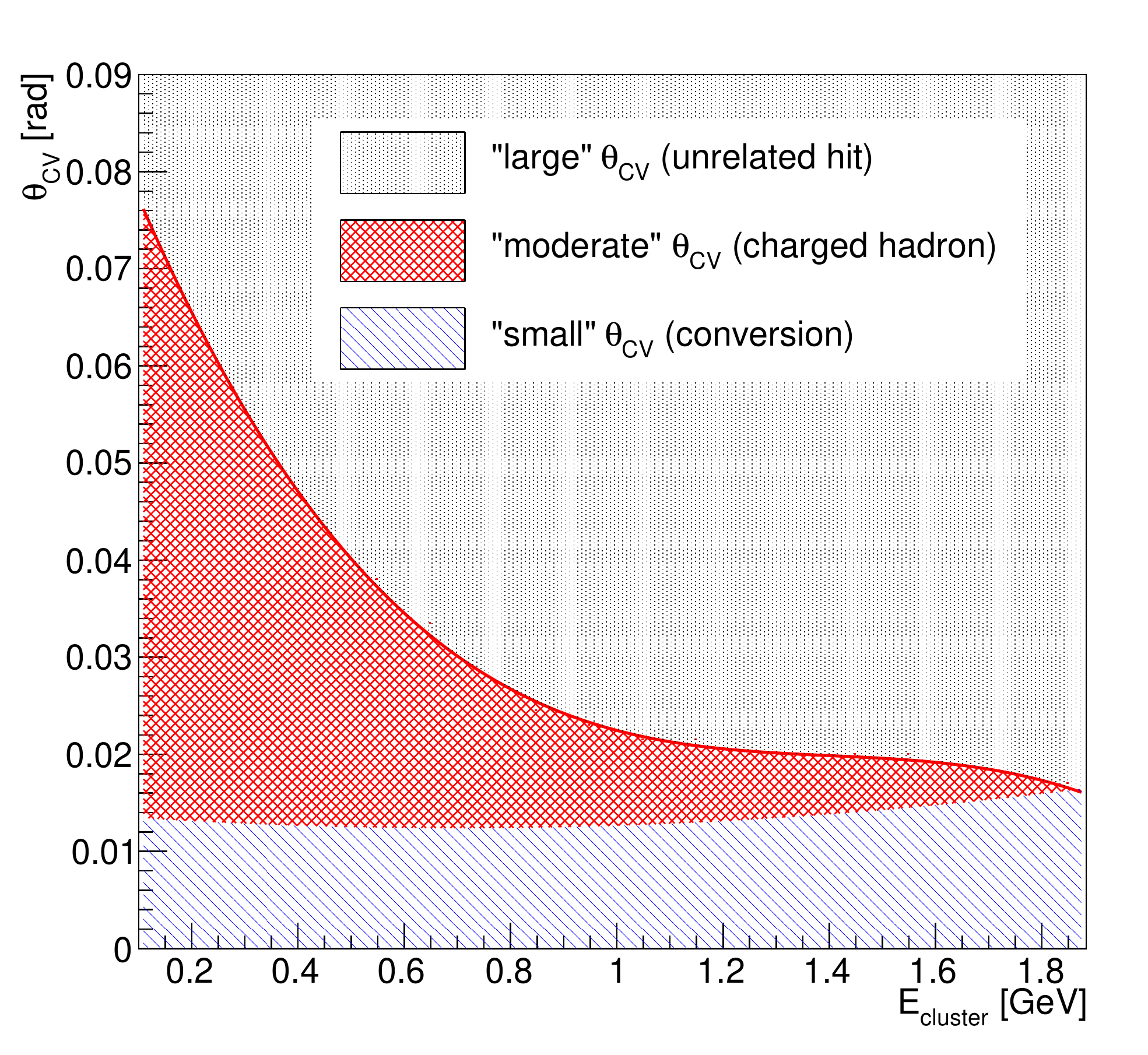}
\caption{(color online)
$\theta_{CV}$ intervals used in the analysis as a function of cluster 
energy in the EMCal PbSc.  Clusters in the red cross-hatched region are 
excluded from the analysis.  For $E_{\rm{cluster}}>1.9$~GeV, no 
distinction between the regions is possible due to the inverse 
relationship between bend and energy for hadronic tracks.}
\label{f:theta_cv_window}
\end{figure}

\begin{figure}
\includegraphics[width=1.0\linewidth]{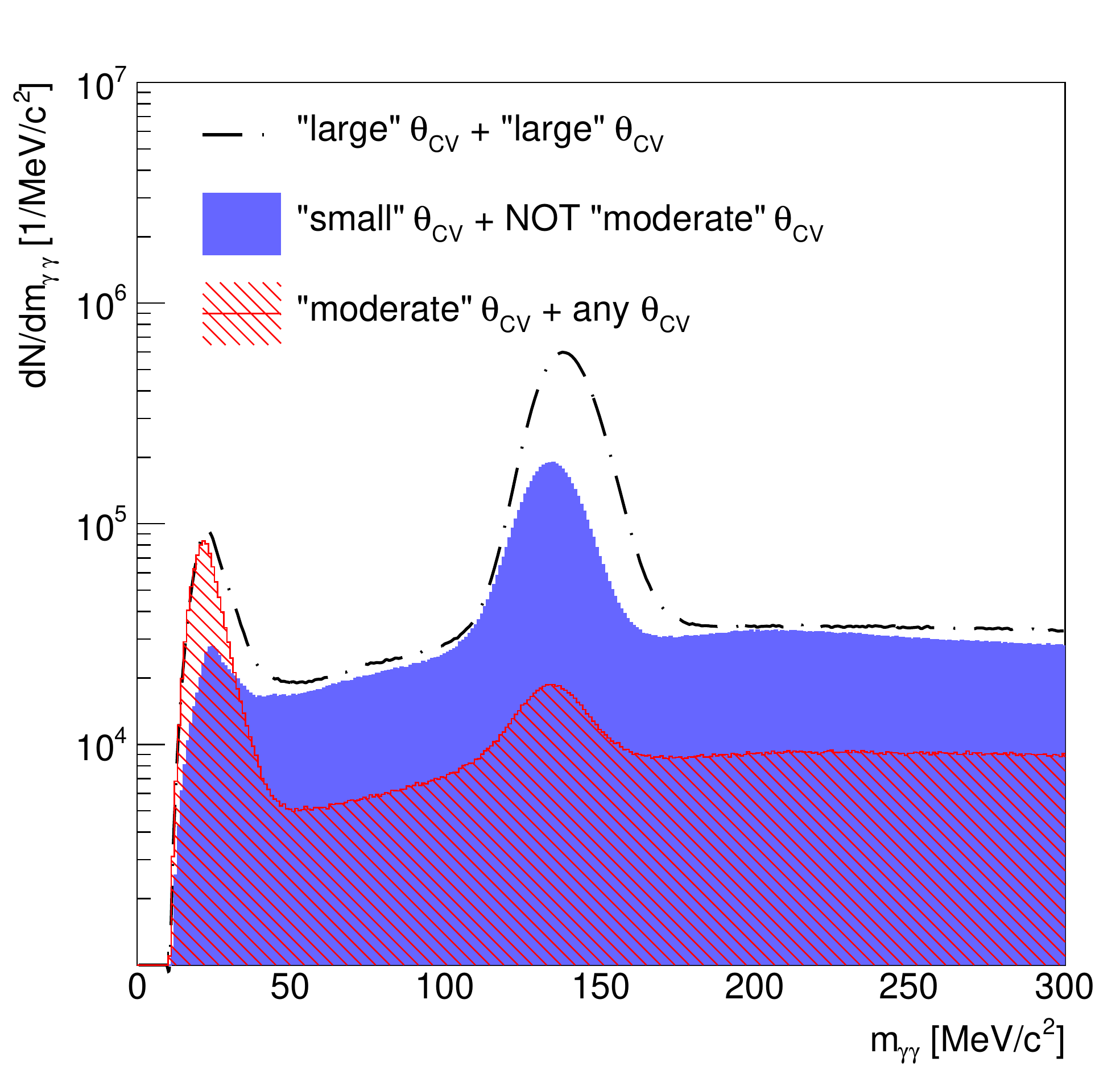}
\caption{(color online)
Yield of cluster pairs in the PbSc with $p_T$ of 2--2.5~\punit for 
different $\theta_{CV}$ requirements as a function of invariant mass 
(calculated assuming both clusters are photons), in the PbSc only and for 
$E_{cluster} < 1.9$~GeV.  The ratio of the ``small'' + ``not moderate'' to 
the ``large'' + ``large'' yield in the \pz mass-peak region is consistent 
with the material budget of $\approx 10\%$ fractional radiation length in 
the magnetic-field region before the PC3.}
\label{f:theta_cv_yields}
\end{figure}

After applying all other cluster cuts, each reconstructed pair invariant 
mass was assigned to the (energy, $\theta_{CV}$) bin of both of its 
clusters, and a $\theta_{CV}$ interval was chosen as a function of cluster 
energy such that the exclusion of clusters in this interval minimized the 
statistical uncertainty on $A_{LL}^{\pz}$ after background \all 
subtraction.  The resulting $\theta_{CV}$ intervals are shown in 
Fig.~\ref{f:theta_cv_window} for clusters in the PbSc with energies below 
$1.9$~GeV, above which the deflection of charged hadrons in the magnetic 
field becomes too small to make a clear separation in $\theta_{CV}$.  Due 
to the decreased response of the PbGl to hadrons, no additional benefit 
for the charge veto cut on top of the other cuts was found and the charge 
veto cut was not applied.  In contrast, when selecting on PbSc clusters 
with energy $< 1.9$~GeV, the charge veto cut improved the statistical 
uncertainty on $A_{LL}^{\pz}$ in the 1--1.5~\punit $p_T$ bin by $5\%$ 
when applied on top of all other cluster cuts.

The invariant mass distribution near the \pz mass peak reconstructed using 
clusters in the PbSc is shown in Fig.~\ref{f:theta_cv_yields} for 
different $\theta_{CV}$ requirements.  It is clear that the signal to 
background ratio for the \pz meson is significantly smaller for clusters 
with a moderate $\theta_{CV}$, due to hadron contamination in the photon 
candidates.  The sample with one small $\theta_{CV}$ cluster is dominated 
by conversions, and some energy is lost in this process, causing the \pz 
mass peak to reconstruct at slightly lower mass.  The effect of this mass 
shift was studied and found to have a negligible impact on the final 
asymmetries.

\subsubsection{Time of Flight Cut}
\label{ss:tof_cut}

A particular hardware-based effect that became apparent with increases in 
the instantaneous luminosities delivered to the experiments in 2009 
involved the readout electronics for the EMCal.  When the trigger fires, 
the signal in each EMCal tower is compared with an analog-buffered value 
from 424 ns, or four crossings, earlier.  Due to the long decay time of an 
EMCal signal, any energy deposit occurring in the three previous crossings 
is read out.  Pileup is negligible due to the fine lateral segmentation of 
the EMCal, so only the combinatorial background is affected.  In the 2009 
run, the likelihood for a collision in at least one of three previous 
crossings was significant at about $22\%$.  One cut in particular that can 
reduce this effect is the ToF cut.

The ToF for a given EMCal cluster is given relative to $t_0$, the initial 
time of the collision as measured by the BBC.  Photon candidates in this 
analysis are required to reach the EMCal within $^{+8}_{-6}$~ns of the 
expected ToF for a photon, which removes low energy hadrons and other out 
of time clusters but also reduces the contribution of clusters from 
previous crossings.  Although the circular buffering in the EMCal readout 
makes the ToF measurement insensitive to timing offsets that are multiples 
of the beam-crossing period, the fact that different crossings have 
independent $t_0$ effectively smears the ToF distribution.  This is the 
dominant effect in increasing the likelihood of previous-crossing clusters 
to have a ToF outside the cut window.

This background can be studied in more detail by analyzing specific sets 
of crossings that follow one- or two-bunch empty crossings and therefore 
contain a smaller number of previous-crossing clusters.  We define the 
following crossing selections for study based on the number of previous 
crossings that can contribute clusters given a four-crossing (current plus 
three) memory:

\begin{itemize}
\item $+0$: The three previous crossings are empty
\item $+1$: One of the three previous crossings is filled
\item $+2$: Two of the three previous crossings are filled
\item $+3$: All three previous crossings are filled.
\item $+3b$: Same as $+3$ but spaced further from empty crossings. 
\end{itemize}

\begin{figure}
\includegraphics[width=1.0\linewidth]{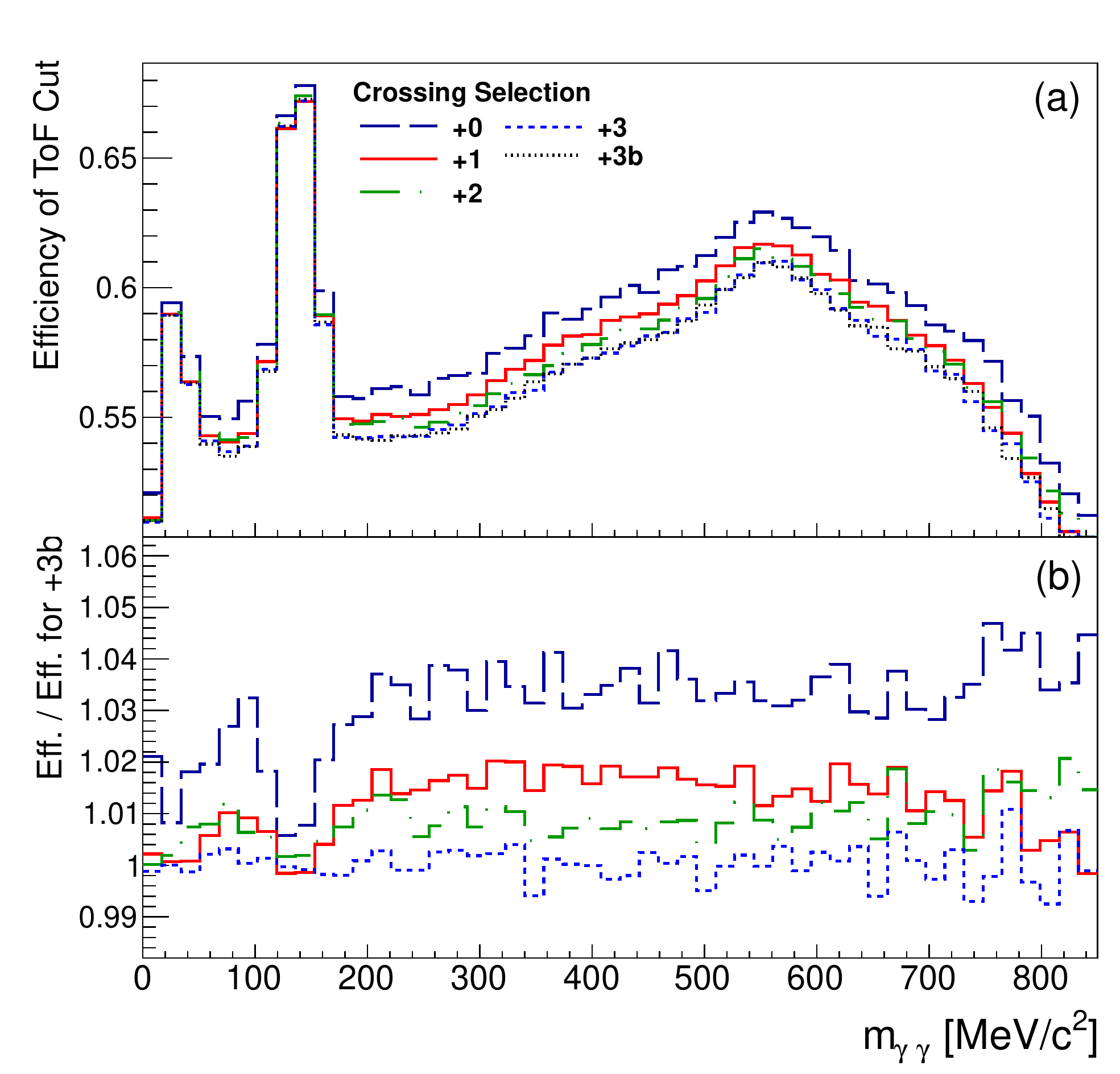} 
\caption{(color online)
(a) Efficiency of ToF cut, with the minimum energy, trigger, and offline 
z-vertex cuts already applied, for different crossing selections defined 
in the text, and for a $p_T$ range of 2--4~\punit.  The energy asymmetry 
cut has not been applied here, and the decreased efficiency in the \et 
mass region is due to the larger background fraction. (b) Ratio of the 
histograms in (a) to the histogram with crossing selection +3b.}
\label{f:emcalmemory_with_tof_cut}
\end{figure}

Figure \ref{f:emcalmemory_with_tof_cut} shows the efficiency of (fraction 
of events passing) the ToF cut on the various selections.  The efficiency 
decreases as the selection moves away from the empty crossings and the 
previous-crossing cluster background increases, indicating that the ToF 
cut is more effective at removing this specific type of background than 
the total background.  Also, from selection $+0$ to $+3$, the relative 
efficiency in the \pz peak region decreases by about $0.5\%$ compared to a 
decrease of roughly $3\%$ in the high mass background efficiency.  The 
smaller change for the peak region is due to the trigger cut (see next 
section) removing true mesons from previous crossings.  As expected, there 
is no significant change in cut efficiency between selections $+3$ and 
$+3b$ since the buffer encompasses only three previous crossings.

\subsection{\pz and \et Selection}
\label{ss:meson_sel}

From the photon candidates surviving the cuts discussed above, all 
combinatorial pairings are reconstructed using the relation for a decay 
into two massless photons,
\begin{equation}
\label{e:inv_mass}
m^2_{\gamma\gamma} \equiv 2E_1E_2(1-cos\theta),
\end{equation}
where $E_1$ and $E_2$ are the energies of the two clusters and $\theta$ is 
the angle between the two vectors from the decay vertex (assumed here 
equal to the collision vertex, which has a negligible impact on 
resolution) to the EMCal clusters.

An additional cut is applied to the photon pairs to ensure that they 
triggered the event, so as to not introduce a bias towards higher 
multiplicity events or convolute the \pz- or \et-meson asymmetry with that 
of a different trigger particle.  All trigger tiles overlapping a 
$12\times12$ tower region ($\Delta \eta \sim 0.1$, $\Delta \phi \sim 0.1$ 
rad.) are read out as one ``supermodule,'' which is the smallest 
segmentation in the recorded trigger information.  We require that the 
central tower of the higher energy photon candidate cluster be located 
within a supermodule firing the trigger.  This also effectively guarantees 
that the cluster comes from the current, and not a previous, crossing.

To further reduce the background for the \et, an energy asymmetry cut is 
applied to exclude cluster pairs, where
\begin{equation}
\frac{|E_1 - E_2|}{E_1 + E_2} \geq a.
\label{e:e_asymm}
\end{equation}
For the \et analysis, a value $a = 0.7$ was used, which optimized the 
uncertainty on $A_{LL}^{\eta}$.  The application of this cut in addition 
to all other cluster and pair cuts improved the uncertainty by about 
$50\%$ in the 2--3~\punit $p_T$ bin and about $7\%$ in the 3--4~\punit 
bin.  For the \pzall analysis, the energy asymmetry cut was not used, 
since its application results in a large uncertainty \emph{increase} in 
each \pt bin, owing to the fact that the effect of the energy asymmetry 
cut on signal and background is comparable and the signal to background 
ratio is much higher for the \pz meson.

\begin{figure}
\includegraphics[width=1.0\linewidth]{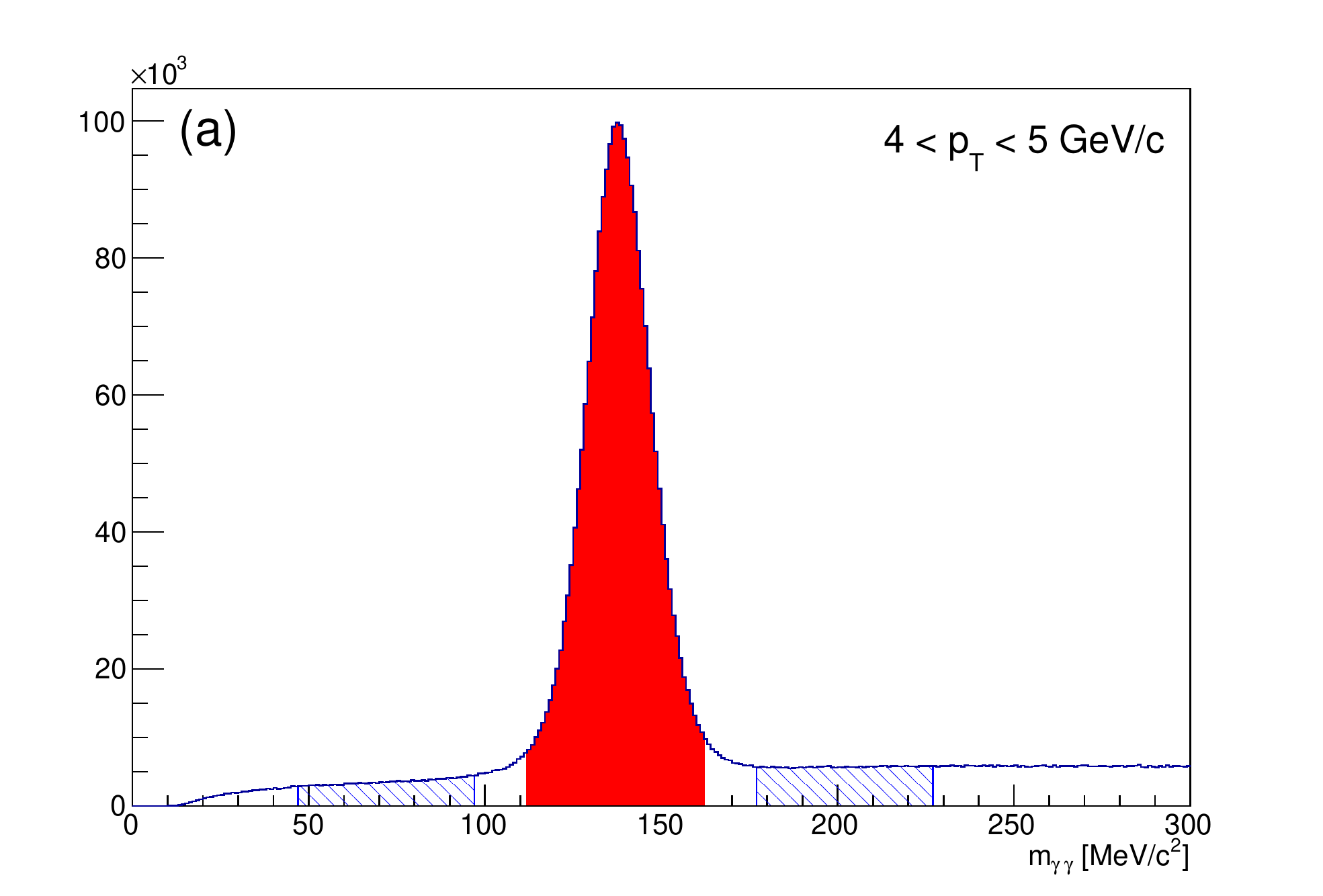} 
\includegraphics[width=1.0\linewidth]{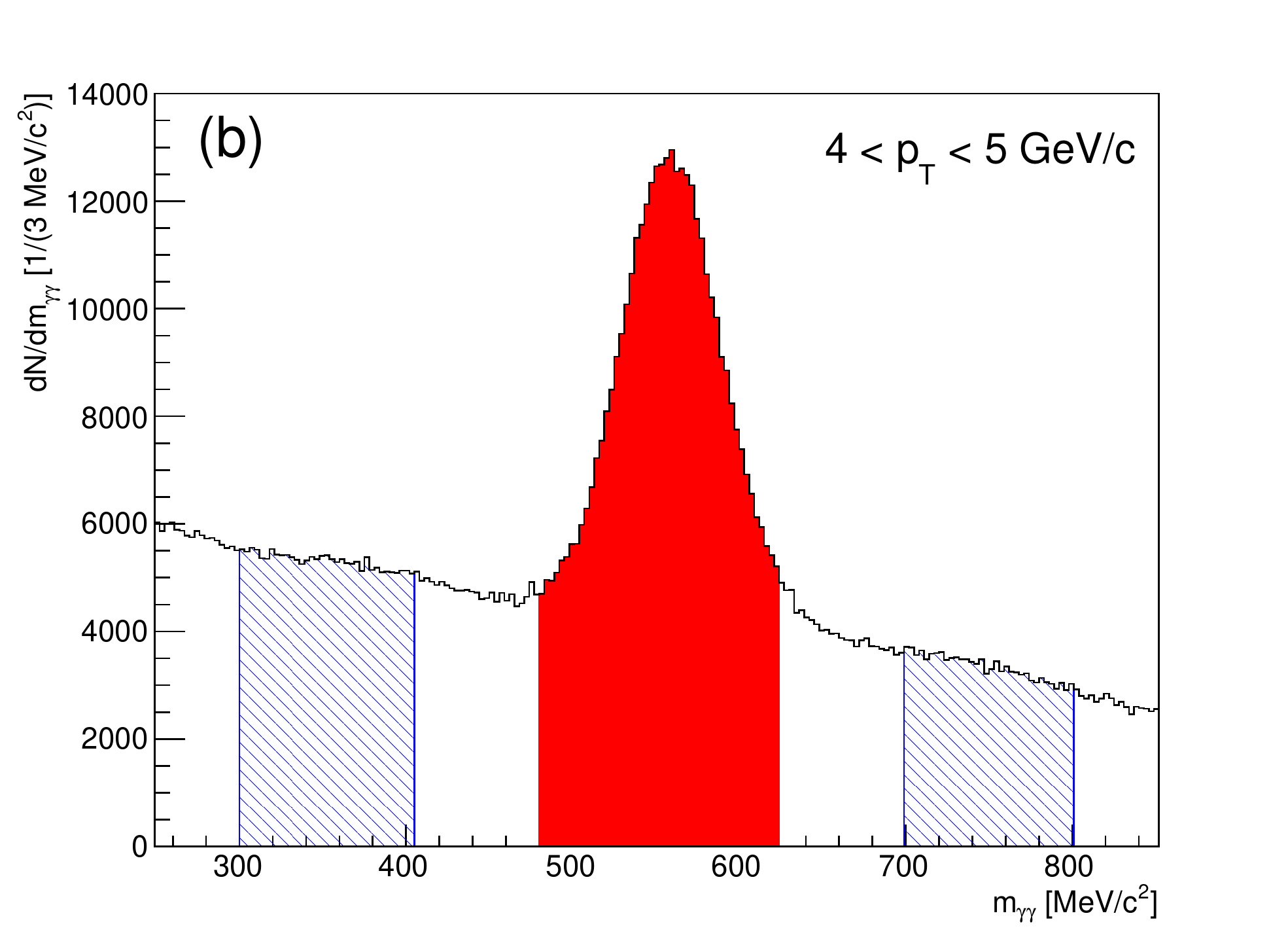} 
\caption{(color online)
(a) Two-photon invariant mass in the region of the \pz mass for the 
$4<\pt<5$~\punit bin with all cuts except the energy asymmetry cut 
applied, as discussed in the text.  (b) Two-photon invariant mass in the 
region of the \et mass with all cuts including the energy asymmetry cut 
applied, as discussed in the text.  In both plots, the red solid region 
shows the signal region and the blue hatched region shows the background 
sidebands used in the asymmetry analysis.}
\label{f:inv_mass_spect}
\end{figure}

The final invariant mass spectra with all cuts applied are shown 
separately for the \pz and \et mesons for a single \pt bin in 
Fig.~\ref{f:inv_mass_spect}.  The signal (solid red) and sideband (hatched 
blue) regions used in the \all analyses are illustrated for each meson.

\subsection{Asymmetry Analysis}

Experimentally, measuring \all as written in Eq.~\ref{eq:allsigma} is not 
feasible due to the sizable systematic uncertainties in any cross section 
measurement, and the small asymmetries expected.  Therefore, \all is 
expressed as
\begin{equation}\all = \frac{1}{P_{B}P_{Y}}\frac{N_{++} - RN_{+-}}{N_{++} + RN_{+-}}
\label{eq:allyields}
\end{equation}
where $N$ is the observable meson yield in the given helicity state and 
$P_{B(Y)}$ is the polarization of the blue (yellow) beam.  $R$ is the 
relative luminosity between helicity states, and is defined as
\begin{equation}
R = \frac{L_{++}}{L_{+-}}\;
\label{eq:rellumi}
\end{equation}
where $L$ is the luminosity sampled in each helicity state.

By writing \all in this way, we are implicitly assuming that all 
acceptance and efficiency corrections are helicity and crossing 
independent.  The detector acceptance and reconstruction efficiencies do 
not change on the scale of hundreds of nanoseconds, which is the typical 
time between helicity flips in RHIC, so these are not an issue.  In the 
case of the trigger efficiency, however, this assumption does not hold due 
to the design of the trigger circuit.  As discussed in 
Sec.~\ref{s:phenix}, the even and odd crossings have different trigger 
circuits with different effective trigger thresholds.  Therefore, the 
analysis is done separately for odd and even crossings for $\pt<7$~\punit.  
Above this $p_T$, the triggers are maximally efficient and there is no 
observed dependence on the trigger circuit.

Similarly, for $R$, we do not measure the ratio of luminosities recorded 
in each helicity state, but instead the ratio of MB triggered events, 
again assuming that efficiency and acceptance cancel in the ratio.  The 
accuracy of this assumption, as well as the assumption that the MB trigger 
has no inherent asymmetry, are discussed below.  The latter leads to the 
largest systematic uncertainty in the determination of \all.

As seen in Fig.~\ref{f:inv_mass_spect}, the two-photon mass yield in the 
\pz or \et mass-peak region (solid red shading) comprises both signal 
and background.  The asymmetry measured in this region, $A_{LL}^{S+B}$, 
mixes both the signal asymmetry, $A_{LL}^{S}$, and the asymmetry in the 
background component, $A_{LL}^{B}$.  The relationship between these three 
asymmetries in the mass peak region can be written as
\begin{eqnarray}
\label{eq:allbkgdsub}
A_{LL}^{S} &=& \frac{A_{LL}^{S+B} - w_{\rm BG}A_{LL}^{B}}{1-w_{\rm BG}} \\ \nonumber
\Delta A_{LL}^{S} &=& \frac{\sqrt{(\Delta A_{LL}^{S+B})^2 
+ r^2 (\Delta A_{LL}^{B})^2}}{1-r}, 
\end{eqnarray}
where $w_{\rm BG}$ is the background fraction in the peak region.  For the 
\pz meson, we define the peak region as $112<\mgg<162$~\munit, which 
corresponds to roughly $2\sigma$ about the mean of the mass peak at low \pt.  
Similarly, for the \et meson, the peak region is defined as 
$480<\mgg<620$~\munit.  The peak positions do not correspond exactly to 
the known mass values for the mesons due to energy smearing effects in the 
EMCal.

The background fraction $w_{\rm BG}$ is extracted from a fit to the mass 
range near the meson mass peak:  50--300~\munit for the \pz meson, and 
300--800~\munit for the \et meson.  In both cases, the fit function 
comprises a 
Gaussian to describe the mass peak plus a third-order polynomial to 
describe the background.  $w_{\rm BG}$ is defined as the integral of the 
background polynomial in the mass peak range $[m_1,m_2]$ divided by the 
total yield in this same range:
\begin{equation}
w_{\rm BG} = \frac{\int_{m_1}^{m_2}dm(a_0+a_1m+a_2m^2+a_3m^3)/\mbox{bin width}}{\mbox{Yield}_{[m_1,m_2]}}.
\label{eq:r_calc}
\end{equation}
Variations of the initial fit parameters, range, and histogram binning 
showed no significant modification to $w_{\rm BG}$ except in the 
$12-15$~\punit \pt bin, where modifying the binning led to a $2.1\%$ 
change in $A_{LL}^{\pi^0}/\sigma_{A_{LL}^{\pi^0}}$, attributable to the 
difficulty in fitting the low-statistics background in this $p_T$ range.  
Average background fractions for the different \pt bins are listed in 
Table~\ref{t:bgfrac}.

\begin{figure}
\includegraphics[width=1.0\linewidth]{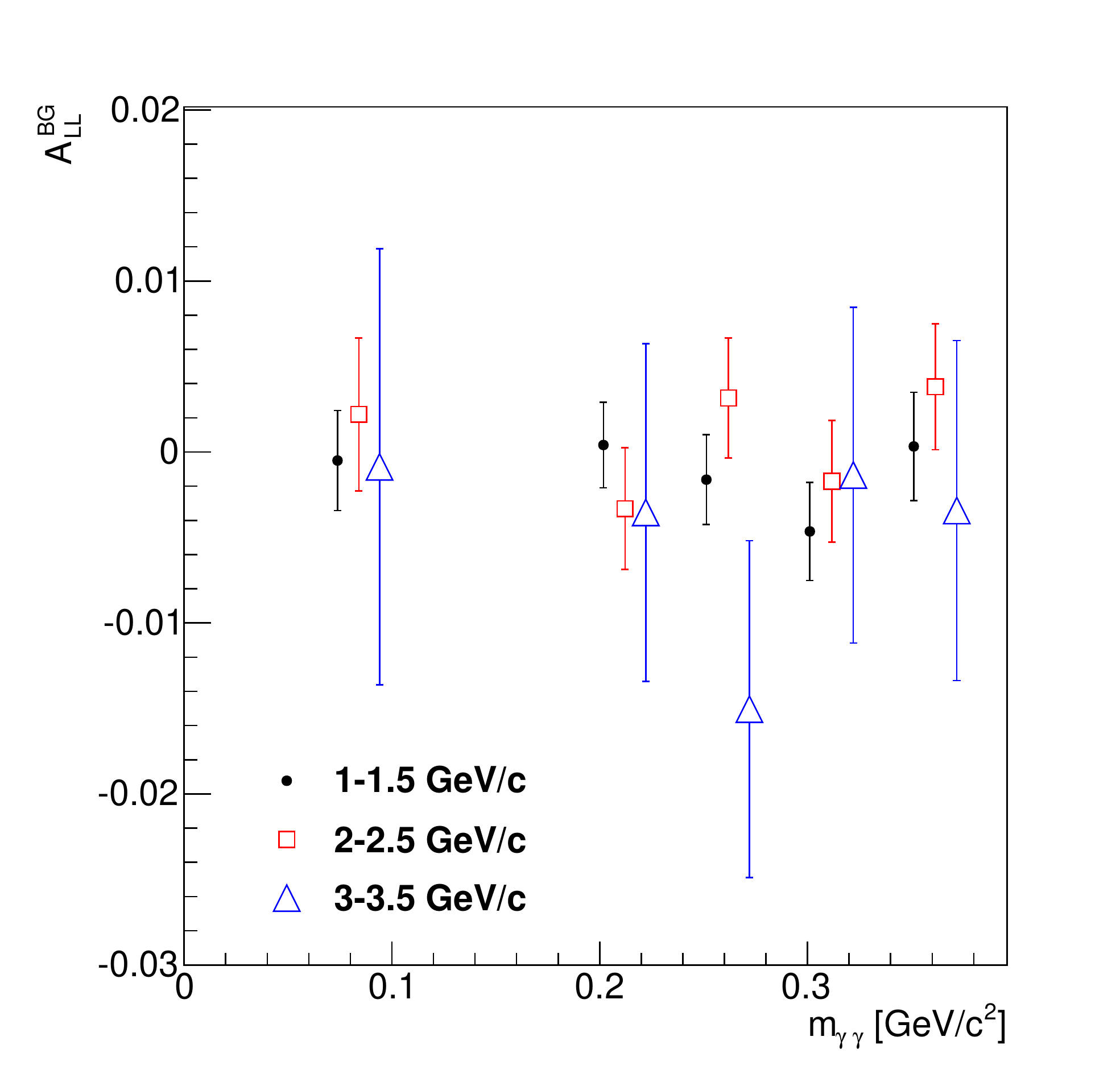}
\caption{(color online)
\all vs. \mgg for the background region near the \pz mass peak for three 
$p_T$ bins: 1--1.5~\punit (black circle), 2--2.5~\punit (red square) and 
3--3.5~\punit (blue triangle), for a single spin pattern in odd 
crossings.  The peak region \all is not shown.  No \mgg dependence is 
found.}
\label{f:allvsmass}
\end{figure}

The background asymmetry in the peak region cannot be directly measured, 
but if the background asymmetry is constant as a function of 
$m_{\gamma\gamma}$, then a measurement in the sideband regions on either 
side of the peak can be used instead.  Figure~\ref{f:allvsmass} shows the 
asymmetry as a function of mass in the background region near the \pz peak 
for several \pt bins.  No indication of a mass dependence in the 
background asymmetry is seen.  However, as discussed below, a small 
systematic uncertainty is evaluated for \pzall to account for any mass 
dependence.  In the case of \etall, any background dependence is 
negligible when considering the limited statistics.  To increase 
statistics, the yields in the sidebands on both sides of the peak region 
are summed to calculate the background asymmetry.  The sideband regions 
are shown in Fig.~\ref{f:inv_mass_spect}, and for the \pz meson are 
defined as $47<\mgg<97$~\munit and $177<\mgg<227$~\munit.  For the \et 
meson, they are $300<\mgg<400$~\munit and $700<\mgg<800$~\munit.

As written in Eq.~\ref{eq:allyields}, \all is calculated for peak and 
background sidebands in each RHIC fill.  Due to the variation in trigger 
electronics discussed above, the analysis is done separately for even and 
odd crossings.  For each of the four spin patterns, $A_{LL}^{S}$ in even 
or odd crossings is calculated using Eq.~\ref{eq:allbkgdsub} with the 
statistically-weighted-average over fills of $A_{LL}^{S+B}$ and 
$A_{LL}^{B}$.  The eight results (four spin patterns for even crossings 
and four spin patterns for odd crossings) are found to be consistent and 
combined to arrive at the final $A_{LL}^{S}$.

\begin{table}[tbp]
\caption{Average background fractions under the $\pi^0$ and $\eta$ peaks, 
$w_{\rm BG}^{\pi^0}$ and $w_{\rm BG}^{\eta}$, in each $p_T$ bin for the 
2009 data.  
In the actual analysis, separate calculations of $w_{\rm BG}$ were used in 
different data subsets (e.g., even and odd crossings).}
\begin{ruledtabular} \begin{tabular}{ccccccc}
& $p_T^{\pi^0}$ bin  
& $w_{\rm BG}^{\pi^0}$ &
& $p_T^{\eta}$ bin  
& $w_{\rm BG}^{\eta}$ & \\
& (GeV/$c$)  & (\%)  & 
& (GeV/$c$)  & (\%)  & \\
\hline
& 1--1.5 & 49 & &  & & \\
& 1.5--2 & 34 & &  & & \\
& 2--2.5 & 23 & &  2--3 & 78  & \\
& 2.5--3 & 17 & &  & & \\
& 3--3.5 & 13 & &  3--4 & 57  & \\
& 3.5--4 & 12 & &  & & \\
& 4--5   & 11 & &  4--5 & 46  & \\
& 5--6   & 11 & &  5--6 & 43  & \\
& 6--7   & 10 & &  6--7 & 43  & \\
& 7--9   & 10 & &  7--9 & 39  & \\
& 9--12  & 9.1 & & & & \\
& 12--15 & 5.5 & & & & \\
\end{tabular} \end{ruledtabular}
\label{t:bgfrac}
\end{table}

\subsection{Systematic Uncertainties}
\label{ss:systematics}

In this section we discuss the systematic uncertainties relevant to the 
$\pi^0$ and $\eta$ analyses, chief among them the uncertainty in the 
determination of relative luminosity.  The various contributions are 
summarized in Table~\ref{t:syst}.

\begin{table}[tbp]
\caption{Summary of systematic uncertainties on \pz and \et \all for the 
2009 data.  The systematics listed as ``\pz only'' were not evaluated for 
the \et asymmetries.}
\begin{ruledtabular} \begin{tabular}{cccc}
 Description
& $\Delta A_{LL}(\mbox{syst})$
& $p_T$
& Note \\ 
& & correlated? & \\
\hline 
 Relative luminosity
& $1.4 \times 10^{-3}$ 
& Yes 
& - \\ 
 Pol. magnitude
& $0.065 \times A_{LL}$
& Yes 
& - \\
 Pol. direction
& $^{+0.026}_{-0.042} \times A_{LL}$
& Yes 
& - \\ 
 $w_{\rm BG}$ determ.
& $<0.01 \times \Delta A_{LL}^{\mbox{stat}}$  
& No 
& \pz only \\ 
 EMCal readout
& $1.6 \times 10^{-4}$ 
& No 
& \pz only, \\
& & \multicolumn{2}{r}{lowest \pt bin} \\
\end{tabular} \end{ruledtabular}
\label{t:syst}
\end{table}

\subsubsection{Relative Luminosity}
\label{ss:rel_lumi}

To account for luminosity differences between same ($++$) and 
opposite ($+-$) helicity crossings, we include a factor $R$ for relative 
luminosity normalization in Eq.~\ref{eq:rellumi}.  Unlike in lepton-proton 
scattering experiments, where QED calculations are precise enough to 
control for spin asymmetries in the extraction of relative luminosity from 
the inclusive DIS cross section, there is no suitable process in 
$\vec{p}+\vec{p}$ that is both high rate and precisely calculable.  For 
absolute luminosity in cross section measurements, we use a machine 
luminosity calculated from beam currents and beam spatial profiles, the 
latter of which are extracted via an experimental technique called a 
Vernier Scan \cite{ref:vern}.  The resulting uncertainty on this machine 
luminosity is too large for use in asymmetry calculations.  However, 
accurate measurements of $R$ can be made using any detector insensitive to 
physics asymmetries.

For our purposes, we use the ratio of two-arm coincidence BBC MB triggers 
with a reconstructed vertex $|z|<30$~cm as $R$:
\begin{equation}
R = \frac{N_{++}^{\rm BBC}}{N_{+-}^{\rm BBC}}.
\label{eq:rellumi_meas}
\end{equation}
However, we should be careful that this $R$ is not biased by sensitivity 
of the BBC to some unmeasured physics asymmetry.  To test for sensitivity 
of the BBC to a double helicity asymmetry, we compare to two-arm 
coincidence ZDC triggered events (also with a reconstructed vertex 
$|z|<30$~cm) via
\begin{eqnarray}
A_{LL}^{R}& \equiv &\frac{1}{P_B P_Y} \frac{r_{++} - r_{+-}}{r_{++} + 
r_{+-}}, \label{eq:zdc_bbc_all} \\
r & \equiv & \frac{N_{ZDC}}{N_{\rm BBC}}. \nonumber
\end{eqnarray}
We take the resulting asymmetry \emph{plus} its statistical uncertainty as 
a systematic uncertainty on our knowledge of the double helicity asymmetry 
of BBC triggered events.  We choose the ZDC for comparison because, in 
addition to having a different geometrical acceptance (see Section 
\ref{s:phenix}), it samples a significantly different class of events than 
the BBC.  The BBC fires predominantly on charged particles and is 
dominated by low-\pt soft physics, whereas the ZDC samples mainly 
diffractive physics and, due to its location behind the accelerator's 
bending magnets, which sweep away most charged particles, fires on 
neutrons, photons, and hadronic showers from scattered protons interacting 
with the machine elements.  The asymmetries in the different physics 
sampled by the ZDC and the BBC cannot be directly calculated.  However, 
comparing these two detectors with different physics sensitivities 
increases the likelihood that any nonzero asymmetries would be apparent.

\begin{table}[htbp]
\caption{Measured value of $A_{LL}^{R}$ in $\sqrt{s} = 200$~GeV 
$\vec{p}+\vec{p}$ running in the given years.  $A_{LL}^{R}$ \emph{plus} 
its uncertainty is used as the total shift uncertainty for any physics 
asymmetry result using the BBC as a relative luminosity monitor.  The 
run-year-correlated part of the uncertainty is taken to equal the maximal 
overlap in $A_{LL}^R$ across years: $0.42 \times 10^{-3}$.  The remaining 
part of each year's $A_{LL}^R$ plus its statistical uncertainty is taken 
as a run-year uncorrelated part.}
\begin{ruledtabular} \begin{tabular}{ccc}
  Data Year
& $A_{LL}^R$
& $\Delta A_{LL}^R(\mbox{stat+syst})$ \\ 
& $\times 10^{-3}$
& $\times 10^{-3}$ \\
\hline
  2005 
& $0.42$ 
& $0.23$ \\
  2006 
& $0.49$ 
& $0.25$ \\
  2009 
& $1.18$
& $0.21$  \\
\end{tabular} \end{ruledtabular}
\label{t:RL_years}
\end{table}

\begin{figure}[htb]
\includegraphics[width=1.0\linewidth]{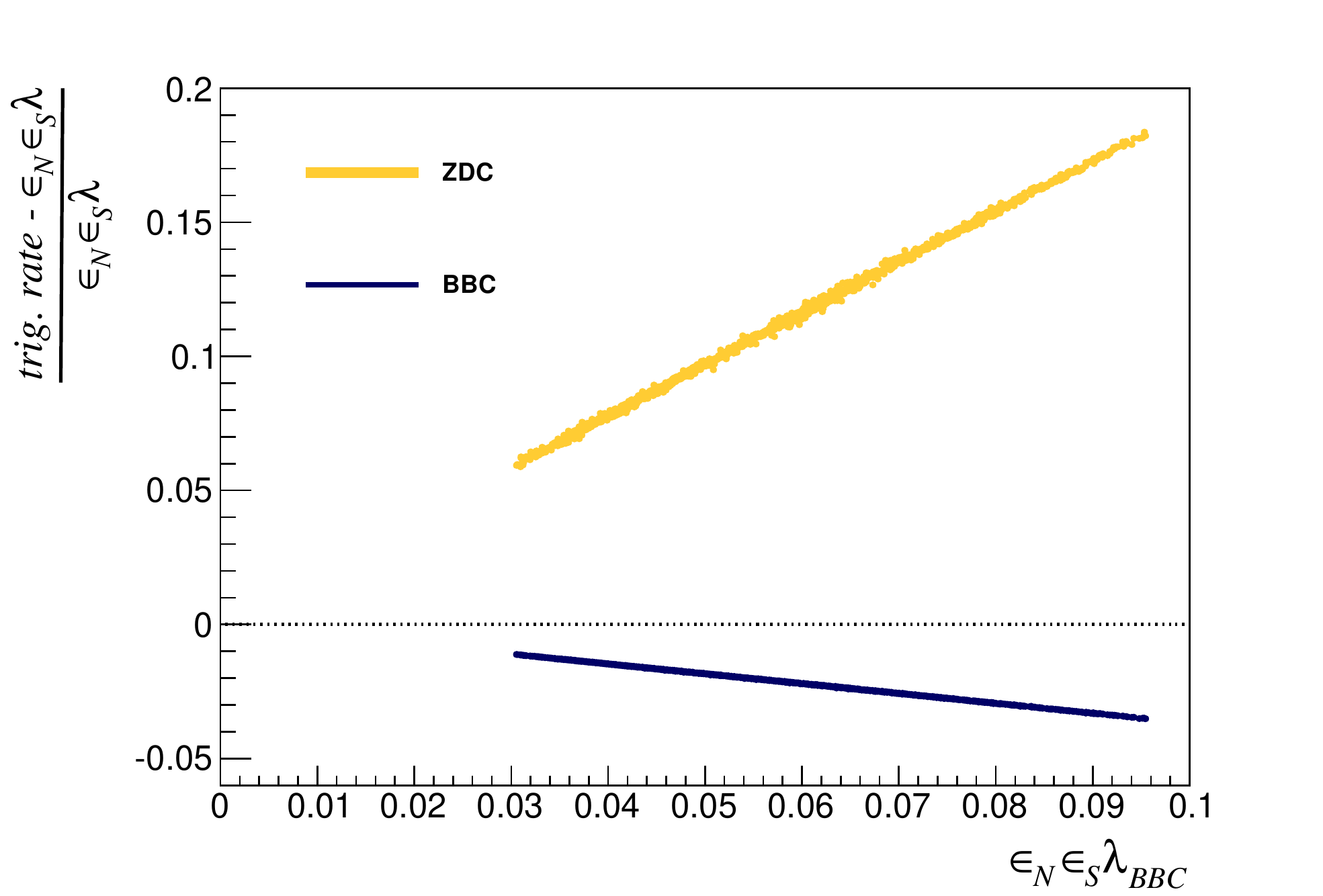}
\caption{(color online)
Relative difference between the measured trigger rate and the quantity in 
Eq.~\ref{eq:eelambda} plotted for all beam crossings in a fraction of the 
present dataset.}
\label{f:under_overcounting}
\end{figure}

Table \ref{t:RL_years} lists the measured asymmetries for three years of 
longitudinally polarized $\vec{p}+\vec{p}$ running at RHIC.  For each 
measurement, a crossing-to-crossing correction for smearing due to the 
$\sim 30$ cm online position resolution of the ZDC was applied but found 
to have little effect on the central $A_{LL}^R$ value or its total 
uncertainty.  Given that $A_{LL}^R$ is significantly higher for the 
present (2009) data, an additional cross-check was performed there, 
motivated by the increased instantaneous luminosity delivered in 2009: the 
calculation of $A_{LL}^R$ using an alternate definition for the 
luminosities sampled by the BBC and ZDC.  The issue is that for any simple 
trigger that returns only one bit of information (yes or no), the ratio of 
triggered events to total $\vec{p}+\vec{p}$ collisions tends to decrease 
with rate as multiple collisions in a single crossing become more common.  
For the BBC, which, accounting for acceptance and efficiency, has a $55\% 
\pm 5\%$ chance to detect a single inelastic $p$$+$$p$ collision, this was the 
dominant effect in the 2009 run.  The ZDC has a much lower efficiency, and 
here the dominant rate effect was instead the increased likelihood of 
coincidence for unrelated background events in the two arms, which lead to 
an increased over-counting of the $\vec{p}+\vec{p}$ collisions.  Using a 
set of scaler boards that were under commissioning during (and thus not 
available over the entirety of) the 2009 run, correlations $N_{OR}$, 
$N_N$, and $N_S$ between single- and double-arm hits were counted in each 
crossing and used to calculate the quantity
\begin{eqnarray}
\epsilon_N \epsilon_S \lambda &=& ln(1-\frac{N_{OR}}{N_{clock}}) \label{eq:eelambda} \\
&-& ln(1-\frac{N_{S}}{N_{clock}}) \nonumber \\
&-& ln(1-\frac{N_{N}}{N_{clock}}), \nonumber
\end{eqnarray}
where $\lambda$ is the average number of events per bunch intersection 
capable of triggering both arms of the detector, and $\epsilon_N$, 
$\epsilon_S$ are factors for the efficiency $\times$ acceptance of the 
arms for these events \cite{ref:myRAPR}.  Because the $z$-vertex cannot be 
reconstructed if only one arm is triggered, this quantity necessarily 
covers the entire $z$-vertex range (the typical collision distribution in 
2009 running had width $\sigma_z \approx 45$~cm).  The advantages of this 
quantity are that it does not under count multiple collisions, and events 
that are not capable of triggering both arms of the detector (such as 
random noise or single-diffractive collisions) are removed analytically.  
The relative difference between $\epsilon_N \epsilon_S \lambda$ and 
trigger rate for the two detectors is shown in 
Fig~\ref{f:under_overcounting}.

The resulting values $\epsilon_N \epsilon_S \lambda$ for the BBC and ZDC 
were used in Eq.~\ref{eq:zdc_bbc_all} with
\begin{equation}
r \equiv \frac{\epsilon_N \epsilon_S \lambda_{ZDC}}{\epsilon_N \epsilon_S \lambda_{\rm BBC}},
\label{eq:eelambda_zdc_bbc_all}
\end{equation}
and the resulting $A_{LL}^R$ was consistent with using the coincidence 
determination in Eq.~\ref{eq:zdc_bbc_all}, and thus the increased 
$A_{LL}^R$ in 2009 over previous years could not be attributed to 
increases in instantaneous luminosity.  The coincidence determination 
yielded the quoted systematic uncertainty
\begin{equation}
A_{LL}^{R} + \delta A_{LL}^{R} = 1.2 + 0.2 \times 10^{-3} = 1.4 \times 10^{-3},
\end{equation}
which is fully correlated across $p_T$ and between the \pz and \et 
results.

\subsubsection{Background Fraction Determination}

Another source of systematic uncertainty arises from the extraction of 
background fractions for the $\pi^0$ and $\eta$ mass peak regions directly 
from the data.  In particular, the background fraction under the peak 
regions is calculated from the result of an empirical fit to the diphoton 
invariant mass spectrum as in Eq.~\ref{eq:r_calc}.

Since the overall normalization is not fixed in the fit and the Gaussian 
part is not used in the calculation, the determination of $w_{\rm BG}$ is 
not particularly sensitive to the shape assumption for the $\pi^0$ mass peak.  
Still, to check for any systematic effect, the $\pi^0$ analysis was rerun 
with the bin width doubled in all invariant mass histograms, which has 
more impact on the resolution of the sharp peak than the relatively flat 
background.  The final $A_{LL}^{\pi^0}$ results changed by less than $1\%$ 
of the statistical uncertainty in all but the $12-15$~GeV/$c$ $p_T$ bin, 
where the change was $2.1\%$.

\subsubsection{Event Overlap in EMCal Readout}

As discussed above in Section \ref{ss:cuts}, readout of the EMCal includes 
clusters from any of the three previous crossings.  The trigger 
requirement ensures that one photon of each pair is in the correct 
crossing, which ensures that true \pz and \et mesons are reconstructed 
from the correct crossing.  However, the combinatorial background may mix 
clusters from previous crossings (with a different helicity combination) 
with clusters from the correct crossing.  The yield of this helicity-mixed 
background depends on the luminosity of previous crossings, and differs 
significantly for crossings following empty crossings.

To test for any impact of this effect on the background asymmetry, \all 
was calculated with a reduced set of cuts using Eq.~\ref{eq:allbkgdsub} 
for the four different spin patterns in RHIC.  Differences were seen in 
the background asymmetries for the different spin patterns, particularly 
at low \pt.  An \mgg dependence in the spin pattern dependent asymmetries 
was also visible.  These effects were mitigated by the full set of cluster 
cuts, including the ToF cut described in Section~\ref{ss:tof_cut}, which 
is more effective than the other cuts in targeting previous-crossing 
background.  Additionally, the asymmetries in the two sidebands and across 
higher mass regions were compared to estimate a possible systematic 
uncertainty arising from any remaining effect.  For the $\pi^0$ analysis, 
the systematic uncertainty in the 1.0--1.5~\punit bin was 
$1.6 \times 10^{-4}$, and for all higher \pt bins the uncertainty was less 
than $10^{-4}$, which is negligible compared to the relative luminosity 
systematic uncertainty as well as the statistical uncertainty.

In addition, to avoid the pooling of data with different nonzero 
background asymmetries, data from the four possible spin patterns were 
analyzed separately through the background subtraction step 
(Eq.~\ref{eq:allbkgdsub}), except for in the $\pi^0$ $9-12$ and 
$12-15$~GeV/$c$ $p_T$ bins where, to increase statistics, patterns 
equivalent for a double-helicity asymmetry (i.e. with ``same'' and 
``opposite'' helicity crossings unchanged) were combined.

\subsubsection{Polarization Direction}

\begin{figure*}[h]
\begin{minipage}{0.48\linewidth}
\includegraphics[width=0.99\linewidth]{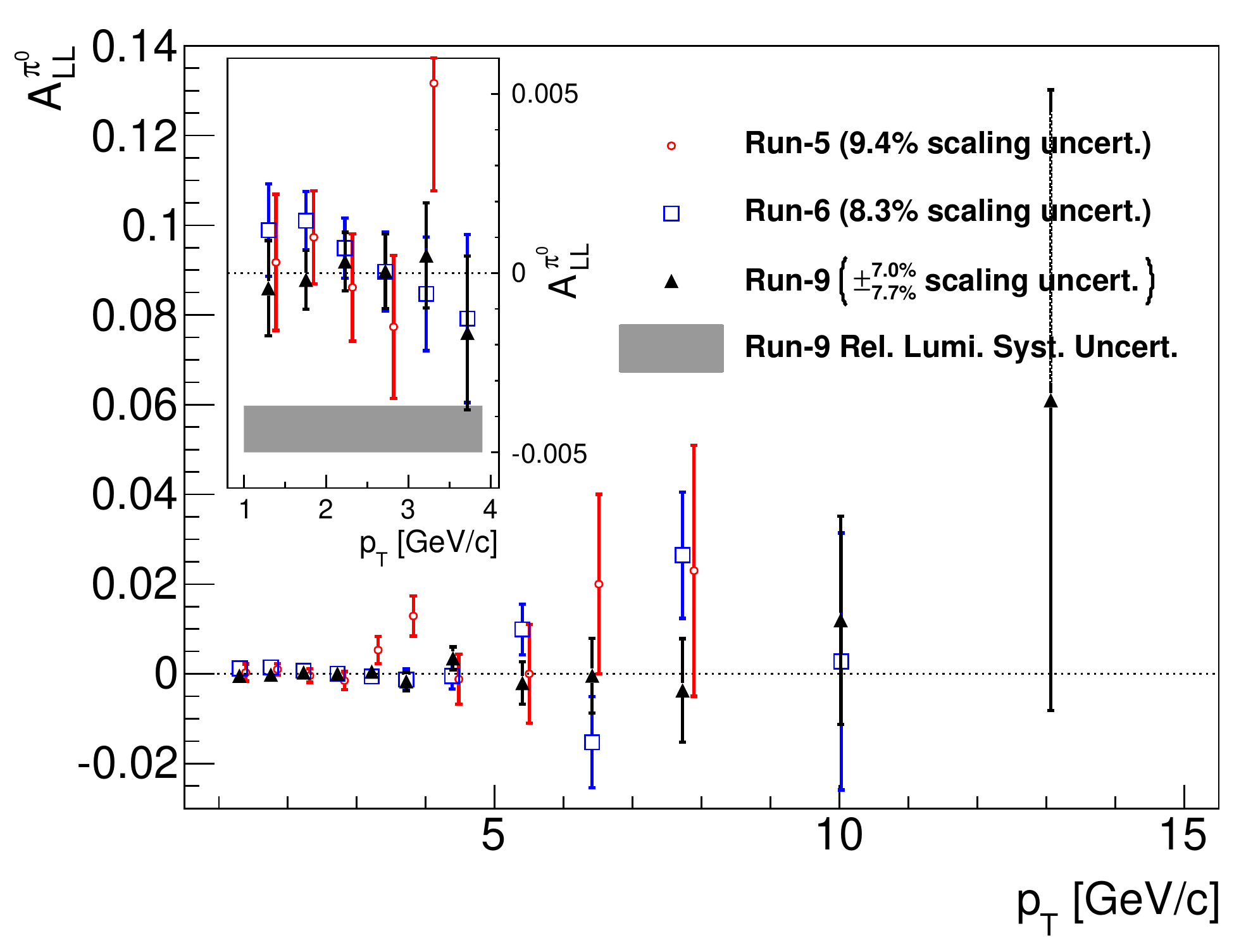}
\caption{(color online)
\all vs.~\pt for \pz mesons for the 2005 (red circle), 2006 (blue square) 
and 2009 (black triangle) PHENIX data sets.  The 2009 relative luminosity 
systematic uncertainty is shown only in the inset, but it applies across 
the entire $p_T$ range.}
\label{f:pi0all569comp}
\end{minipage}
\hspace{0.5cm}
\begin{minipage}{0.48\linewidth}
\vspace{-0.8cm}
\includegraphics[width=0.99\linewidth]{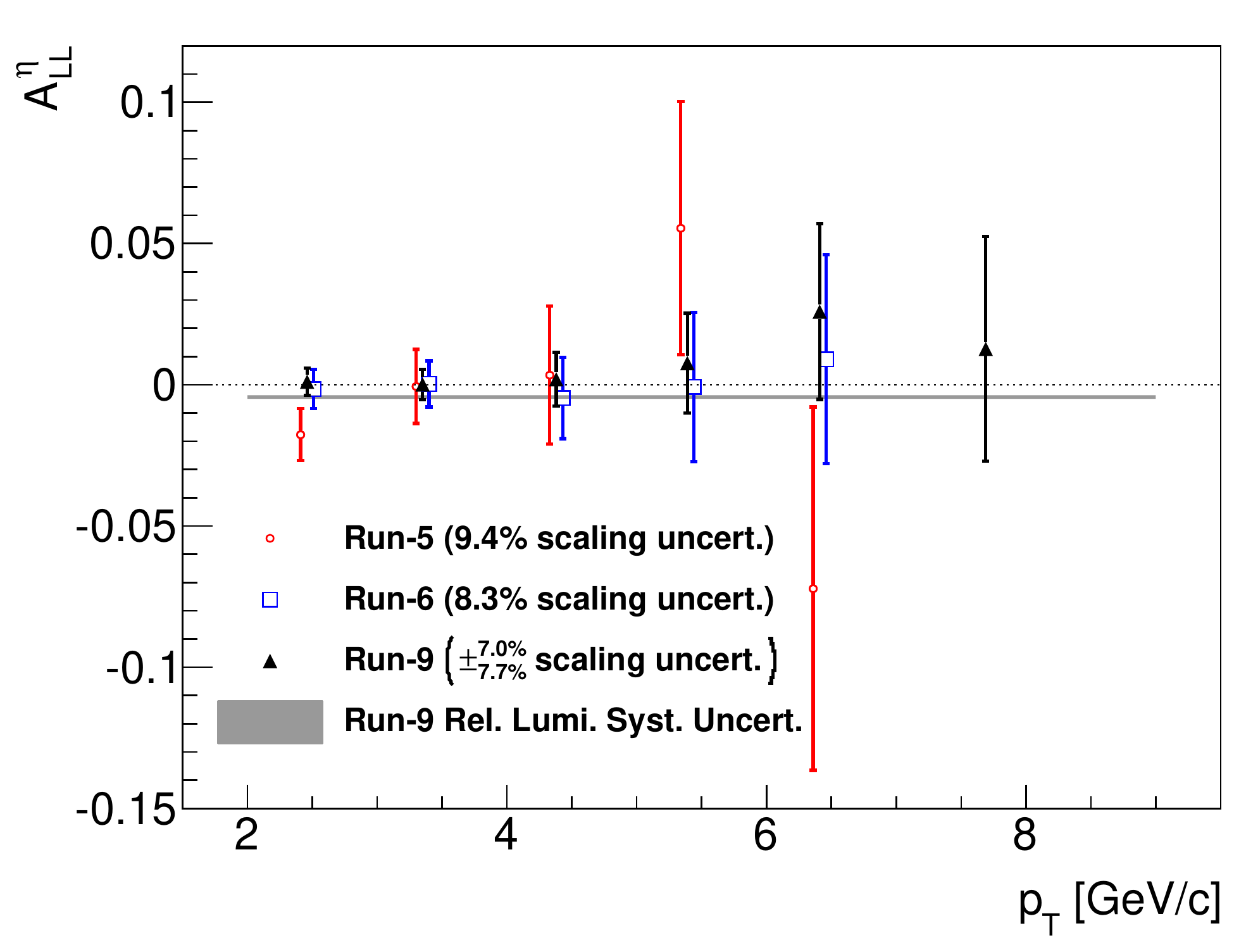}
\caption{(color online)
\all vs.~$p_T$ for \et mesons for the 2005 (red circle), 2006 (blue square) 
and 2009 (black triangle) PHENIX data sets.}
\label{f:etaall569comp}
\end{minipage}
\end{figure*}

Another hardware based uncertainty that has been present in all 
longitudinal RHIC runs is that of the remaining transverse polarization 
component after the beams have passed through the Spin Rotator magnets, 
discussed in detail in Section \ref{s:rhic}. The total resultant scaling 
uncertainty in the longitudinal component of $P_BP_Y$, which applies 
globally to the 2009 dataset, is ($^{+0.026}_{-0.042})$.  This can be 
added in quadrature to the polarization scale uncertainty listed in 
Section~\ref{s:rhic}, and the results of that combination are given in 
Figures~\ref{f:pi0all569comp} and~\ref{f:etaall569comp} as well as 
Tables~\ref{t:pi0_final_datatable} and ~\ref{t:eta_final_datatable}.

\subsubsection{Searches for Additional Systematic Uncertainty Sources}

To test for additional RHIC fill-to-fill uncorrelated systematic 
uncertainties that may have been overlooked, we applied a statistical 
bootstrapping technique to the data.  For each of ten-thousand iterations, 
a separate random spin pattern was chosen for each fill, and all 
quantities were calculated according to this pattern.  This allowed us to 
produce, for the various ``peak'' and ``sideband'' regions, simulated 
distributions for $A_{LL}$, $\delta A_{LL}$, and $\chi^2$ from a fit of 
$A_{LL}$ across RHIC fills.  The result of this test was that the 
uncertainties were being underestimated above $p_T\approx7$~GeV/$c$ for 
the sideband region and overestimated at low $p_T$ for both regions.  The 
sideband region underestimation was traced to low background statistics at 
high $p_T$ resulting in the violation of Gaussian distribution assumptions 
for error propagation.  Since the background fraction $w_{\rm BG}$ is 
small at high $p_T$, this effect is negligible in the final result.  The 
overestimation of uncertainties at low $p_T$ is due to conservative 
calculation of uncertainties in the cases where triggers were scaled to 
match the data acquisition bandwidth. For the $\pi^0$, the largest 
overestimation was about 6\% of the statistical uncertainty, for the 
signal region in the lowest $p_T$ bin.

Measurements of single-spin asymmetries, in which the polarization of one 
beam is summed over, were also performed.  Such asymmetries, if physical, 
would be parity violating.  As expected for a parity-conserving QCD 
process, they were found to be consistent with zero.  Comparisons were 
also made between the two different electromagnetic calorimeter 
technologies.  In these comparisons, both double and single-spin 
asymmetries were measured separately in the PbSc and PbGl, and no 
inconsistency between the two detectors was found.


\begingroup
\squeezetable
\begin{table*}[tbh]
\caption[$\pi^0 A_{LL}$ values from '05, '06, and '09]{$\pi^0$ $A_{LL}$ 
measurements at $\sqrt{s} = 200$~GeV from the 2005, 2006, and 2009 RHIC 
runs, along with statistical uncertainties.  The systematic uncertainties 
for the three years are: relative luminosity (shift uncertainty): $6.5 
\times 10^{-4}$, $7.5 \times 10^{-4}$, and $14.0 \times 10^{-4}$ 
polarization (scale uncertainty): $9.4\%$, $8.3\%$, and 
${}^{+7.0\%}_{-7.7\%}$.}
\begin{ruledtabular} \begin{tabular}{ccccccccccccc}
&& \multicolumn{3}{c}{2005}
&& \multicolumn{3}{c}{2006}
&& \multicolumn{3}{c}{2009} \\ 
  $p_T^{\pi^0}$ bin 
 && $\langle p_T^{\pi^0}\rangle$ 
& $A_{LL}^{\pi^0}$ 
& $\Delta A_{LL}^{\pi^0}$ 
 && $\langle p_T^{\pi^0}\rangle$ 
& $A_{LL}^{\pi^0}$ 
& $\Delta A_{LL}^{\pi^0}$ 
 && $\langle p_T^{\pi^0}\rangle$ 
& $A_{LL}^{\pi^0}$
& $\Delta A_{LL}^{\pi^0}$ \\ 
(GeV/$c$)
&& (GeV/$c$) & $\times10^{-3}$ & $\times10^{-3}$ 
&& (GeV/$c$) & $\times10^{-3}$ & $\times10^{-3}$ 
&& (GeV/$c$) & $\times10^{-3}$ & $\times10^{-3}$ \\
\hline
1--1.5 && 1.29 & 0.3  & 1.9 && 1.30 & 1.2  & 1.3  && 1.30 & -0.4  & 1.3 \\ 
1.5--2 && 1.75 & 1.0  & 1.3 && 1.75 & 1.46 & 0.82 && 1.75 & -0.19 & 0.82 \\ 
2--2.5 && 2.22 & -0.4 & 1.5 && 2.23 & 0.70 & 0.84 && 2.23 & 0.33  & 0.81 \\ 
2.5--3 && 2.72 & -1.5 & 2.0 && 2.72 & 0.0  & 1.1  && 2.72 & 0.1   & 1.0 \\ 
3--3.5 && 3.21 & 5.3  & 3.0 && 3.22 & -0.6 & 1.6  && 3.22 & 0.5   & 1.5 \\ 
3.5--4 && 3.72 & 12.9 & 4.5 && 3.72 & -1.3 & 2.3  && 3.72 & -1.7  & 2.2 \\ 
4--5   && 4.38 & -1.2 & 5.6 && 4.38 & -0.5 & 2.9  && 4.40 & 3.5   & 2.5 \\ 
5--6   && 5.40 & 0    & 11  && 5.40 & 9.9  & 5.7  && 5.40 & -2.1  & 4.7 \\ 
6--7   && 6.41 & 20   & 20  && 6.41 & -15  & 10   && 6.41 & -0.4  & 8.3 \\ 
7--9   && 7.79 & 23   & 28  && 7.74 & 26   & 14   && 7.72 & -4    & 12 \\ 
9--12  && N/A  & N/A  & N/A && 10.0 & 3    & 29   && 10.0 & 12    & 23 \\ 
12--15 && N/A  & N/A  & N/A && N/A  & N/A  & N/A  && 13.1 & 61    & 69 \\ 
\end{tabular} \end{ruledtabular}
\label{t:pi0_final_datatable}
\end{table*}

\begin{table*}[htb]
\caption{$\eta$ $A_{LL}$ measurements at $\sqrt{s} = 200$~GeV from the 
2005, 2006, and 2009 RHIC runs, along with statistical uncertainties.  
The systematic uncertainties for the three years are:  relative 
luminosity (shift uncertainty): $6.5 \times 10^{-4}$, $7.5 \times 
10^{-4}$, and $14.0 \times 10^{-4}$ and polarization (scale uncertainty): 
$9.4\%$, $8.3\%$, and ${}^{+7.0\%}_{-7.7\%}$.}
\begin{ruledtabular} \begin{tabular}{ccccccccccccc}
&& \multicolumn{3}{c}{2005}
&& \multicolumn{3}{c}{2006}
&& \multicolumn{3}{c}{2009} \\ 
  $p_T^{\eta}$ bin 
 && $\langle p_T^{\eta}\rangle$ 
& $A_{LL}^{\eta}$ 
& $\Delta A_{LL}^{\eta}$ 
 && $\langle p_T^{\eta}\rangle$ 
& $A_{LL}^{\eta}$ 
& $\Delta A_{LL}^{\eta}$ 
 && $\langle p_T^{\eta}\rangle$ 
& $A_{LL}^{\eta}$
& $\Delta A_{LL}^{\eta}$ \\ 
(GeV/$c$)
&& (GeV/$c$) & $\times10^{-3}$ & $\times10^{-3}$
&& (GeV/$c$) & $\times10^{-3}$ & $\times10^{-3}$
&& (GeV/$c$) & $\times10^{-3}$ & $\times10^{-3}$ \\
\hline
2--3&&2.41&-17.7&9.2&&2.51&-1.5&6.9&&2.46&1.1&4.9\\ 
3--4&&3.30&-1  &13 &&3.40&0.3 &8.2&&3.35&0.1&5.4\\ 
4--5&&4.33&3   &24 &&4.43&-5  &14 &&4.38&2.0&9.5\\ 
5--6&&5.34&55  &45 &&5.44&-1  &27 &&5.39&8  &18 \\ 
6--7&&6.36&-72 &64 &&6.46&9   &37 &&6.41&26 &31 \\ 
7--9&&N/A &N/A &N/A&&N/A &N/A &N/A&&7.69&13 &40 \\ 
\end{tabular} \end{ruledtabular}
\label{t:eta_final_datatable}
\end{table*}

\endgroup


\begingroup
\squeezetable

\begin{table*}[htb]
\caption{Combined $\pi^0$ $A_{LL}$ values from the PHENIX data sets at 
$\sqrt{s} = 200$~GeV.  Fully \pt correlated systematic uncertainties that 
are considered uncorrelated by run-year are given in the table.  The 
run-year correlated parts of the polarization scale uncertainty, $4.8\%$, 
and the relative luminosity shift uncertainty, $4.2 \times 10^{-4}$, are 
not included.}
\begin{ruledtabular} \begin{tabular}{ccccccc}
 $p_T^{\pi^0}$ bin
& $\langle p_T^{\pi^0}\rangle$
& $A_{LL}^{\pi^0}$
& $A_{LL}^{\pi^0}(\mbox{Stat})$
& $\Delta A_{LL}^{\pi^0}(\mbox{RL Syst})$
& $\Delta A_{LL}^{\pi^0}(\mbox{Pol. Syst})$
& $\Delta A_{LL}^{\pi^0}(\mbox{Tot. Syst})$ \\
 (GeV/$c$)
& (GeV/$c$)
& $\times 10^{-4}$
& $\times 10^{-4}$
& $\times 10^{-4}$
& $\times A_{LL}^{\pi^0}$
& $\times 10^{-4}$ \\
\hline
1--1.5&1.30&5.1 &8.5&3.5&3.4\%&3.6\\ 
1.5--2&1.75&9.6 &5.5&3.3&3.5\%&3.3\\ 
2--2.5&2.23&3.9 &5.8&3.4&3.5\%&3.4\\ 
2.5--3&2.72&-2.3&7.4&3.6&3.4\%&3.7\\ 
3--3.5&3.22&6   &11 &4.0&3.2\%&4.0\\ 
3.5--4&3.72&2   &15 &4.1&3.1\%&4.1\\ 
4--5  &4.39&13  &18 &4.3&3.1\%&4.3\\ 
5--6  &5.40&26  &35 &4.5&3.0\%&4.5\\ 
6--7  &6.41&-39 &61 &4.5&2.9\%&4.6\\ 
7--9  &7.74&96  &85 &4.5&2.9\%&5.3\\ 
9--12 &10.0&80  &180&5.8&3.3\%&6.5\\ 
12--15&13.1&610 &690&10 &3.0\%&21 \\ 
\end{tabular} \end{ruledtabular}
\label{t:pi0_combined_datatable}
\end{table*}

\begin{table*}[htb]
\caption{Combined $\eta$ $A_{LL}$ values from the PHENIX data sets at 
$\sqrt{s} = 200$~GeV.  Fully $p_T$-correlated systematic uncertainties that 
are considered uncorrelated by run-year are given in the table.  The 
run-year correlated parts of the polarization scale uncertainty, $4.8\%$, 
and the relative luminosity shift uncertainty, $4.2 \times 10^{-4}$, are 
not included.}
\begin{ruledtabular} \begin{tabular}{ccccccc}
 $p_T^{\eta}$ bin
& $\langle p_T^{\eta}\rangle$
& $A_{LL}^{\eta}$
& $A_{LL}^{\eta}(\mbox{Stat})$
& $\Delta A_{LL}^{\eta}(\mbox{RL Syst})$
& $\Delta A_{LL}^{\eta}(\mbox{Pol. Syst})$
& $\Delta A_{LL}^{\eta}(\mbox{Tot. Syst})$ \\
 (GeV/$c$)
& (GeV/$c$)
& $\times 10^{-4}$
& $\times 10^{-4}$
& $\times 10^{-4}$
& $\times A_{LL}^{\eta}$
& $\times 10^{-4}$ \\
\hline
2--3  &2.46&-27 &37  &4.5&3.0\%&4.6 \\ 
3--4  &3.35&1   &44  &4.7&2.8\%&4.7 \\ 
4--5  &4.38&2   &77  &4.8&2.8\%&4.8 \\ 
5--6  &5.39&100 &140 &4.8&2.8\%&5.5 \\ 
6--7  &6.41&80  &230 &4.4&3.0\%&5.0 \\ 
7--9  &7.69&130 &410 &10 &3.0\%&11  \\ 
\end{tabular} \end{ruledtabular}
\label{t:eta_combined_datatable}
\end{table*}

\endgroup

\begin{figure*}[htb]
  \includegraphics[width=0.99\linewidth]{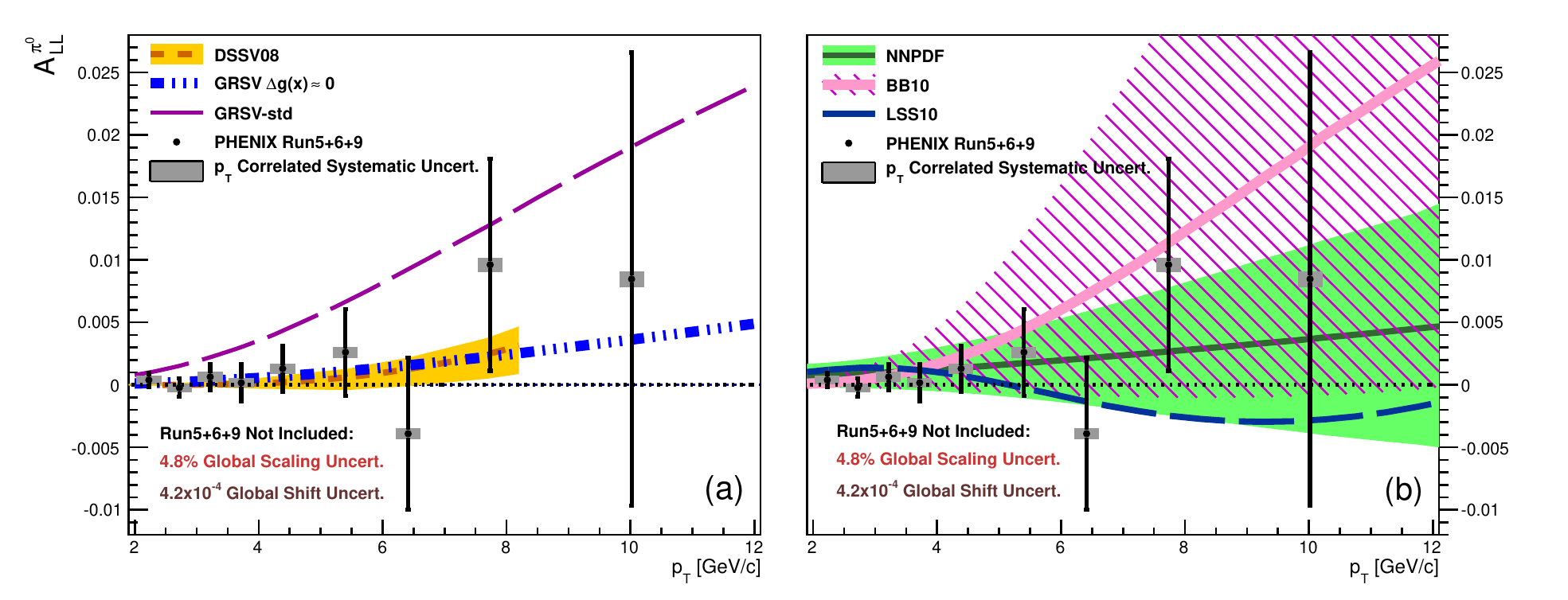}
  \caption[Combined ALL vs. theory]{(color online)
Points are the combined \all vs.~$p_T$ for \pz mesons from 2005 through 2009 
with the statistical uncertainty.  The $p_T$ correlated systematic 
uncertainty given by the gray bands is the result of combining the 
year-to-year uncorrelated parts of the systematic uncertainties on 
relative luminosity and polarization.  The year-to-year correlated parts 
are given in the legend.  Plotted for comparison are several expectations 
based on fits to polarized scattering data, with uncertainties where 
available.  } \label{f:pi0alltheocomp}
\end{figure*}

\clearpage

\section{Results}
\label{s:results}

The \pz and \et \all values as a function of \pt for the 2009 data set are 
shown in Figs.~\ref{f:pi0all569comp} and \ref{f:etaall569comp}, 
respectively, and given in 
Tables~\ref{t:pi0_final_datatable}--\ref{t:eta_combined_datatable}. 
The results are compared with previously published results from 2005 and 
2006~\cite{ref:run5pi0xsect_all,ref:run6pi0_all}, with which they are 
consistent.  The relative luminosity systematic uncertainty for the 2009 
data set is shown only in the inset of Fig.~\ref{f:pi0all569comp} but 
applies to all of the points.  The polarization uncertainties discussed 
above are not shown on the data points but are listed in the legend. The 
results are consistent in all cases.

\section{Discussion}
\label{s:discussion}

\begin{figure}[htb]
\includegraphics[width=1.0\linewidth]{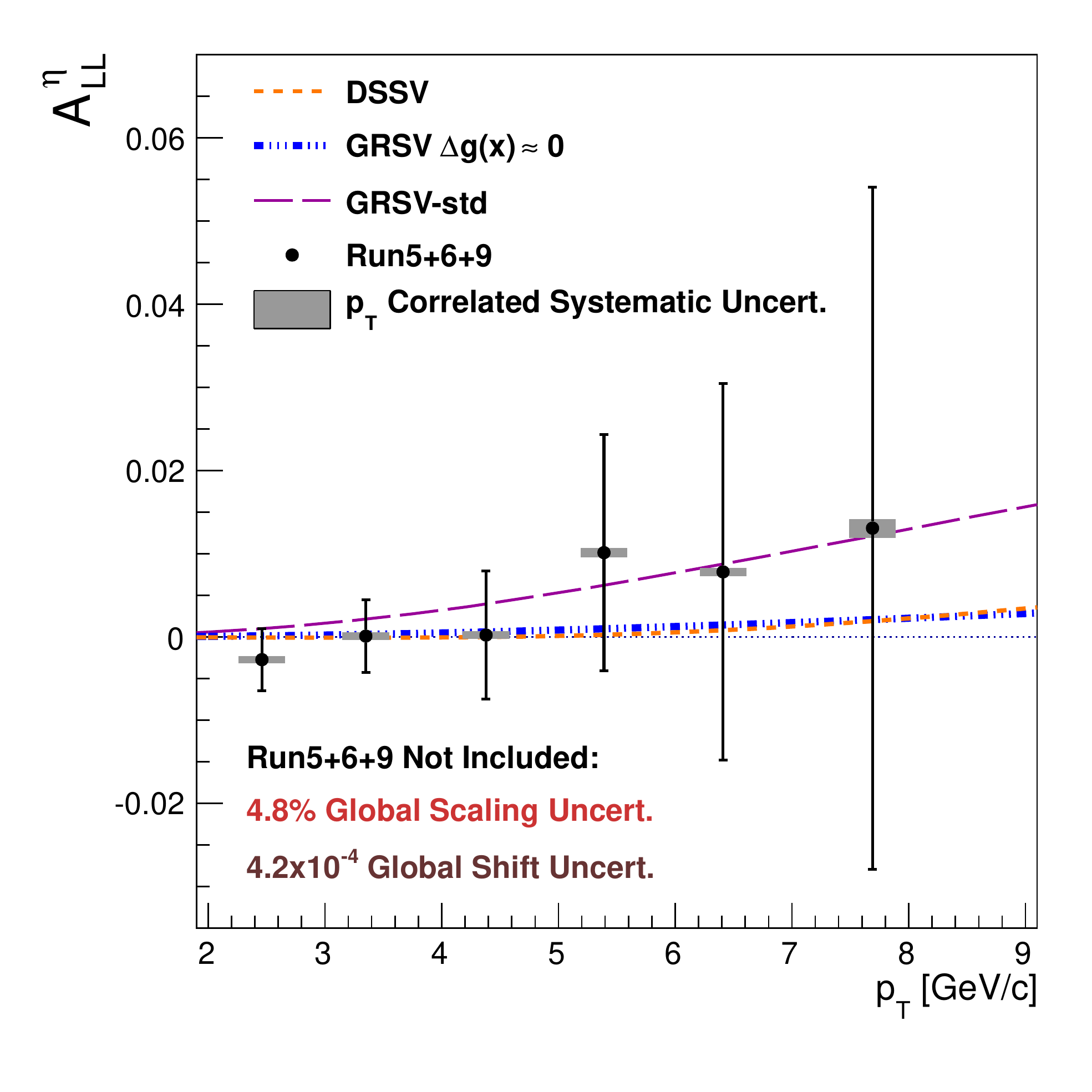}
\caption{(color online)
Points are the combined \all vs.~$p_T$ for \et mesons from 2005 through 2009 
with the statistical uncertainty.  The $p_T$ correlated systematic 
uncertainty given by the gray bands is the result of combining the 
year-to-year uncorrelated parts of the systematic uncertainties on 
relative luminosity and polarization.  The year-to-year correlated parts 
are given in the legend.  Several theoretical expectations based on fits 
to polarized data are also shown, which use results from \cite{ref:eta_FF} 
for the fragmentation functions.}
\label{f:etaalltheocomp}
\end{figure}

\begin{figure*}[htb]
\includegraphics[width=0.48\linewidth]{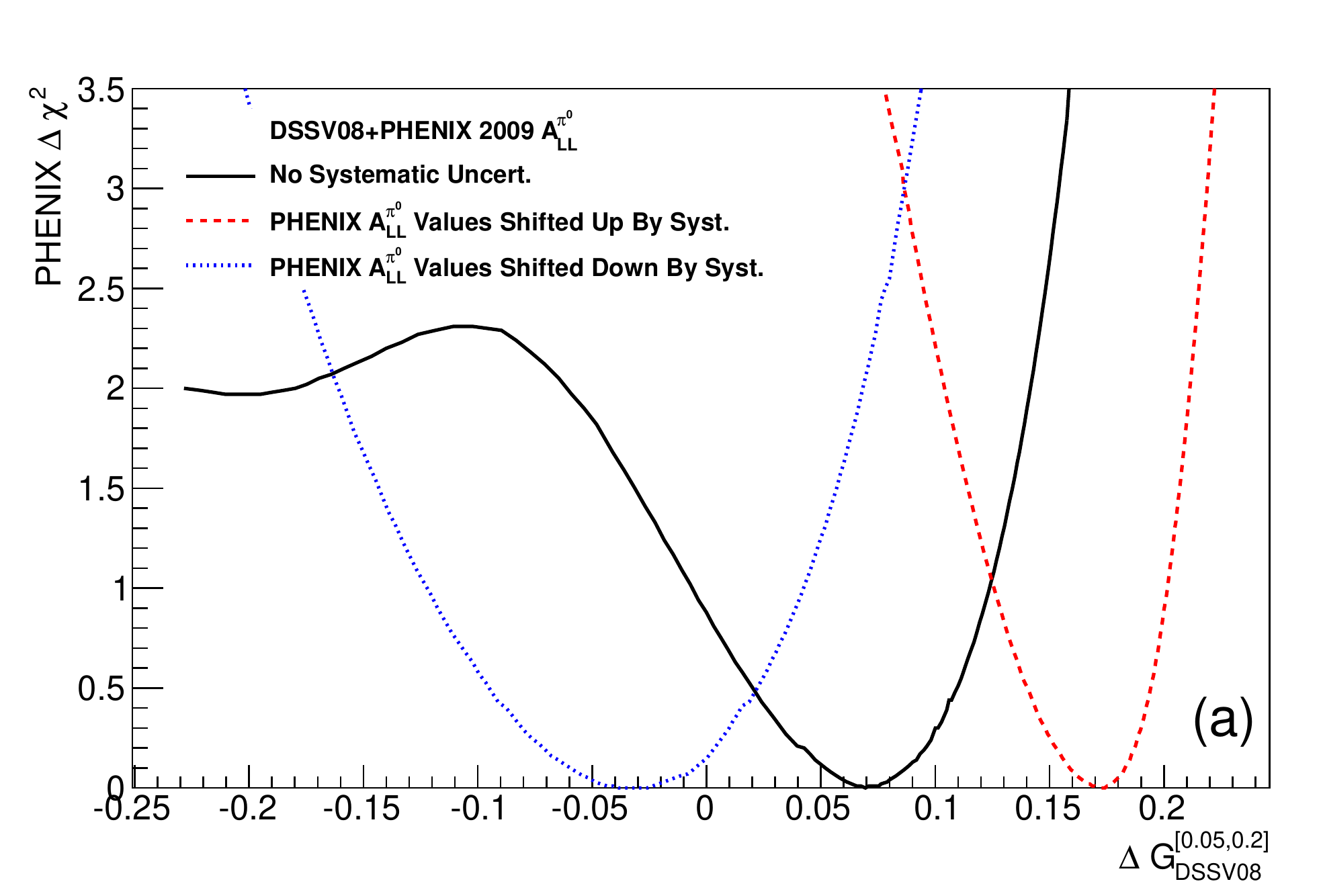}
\includegraphics[width=0.48\linewidth]{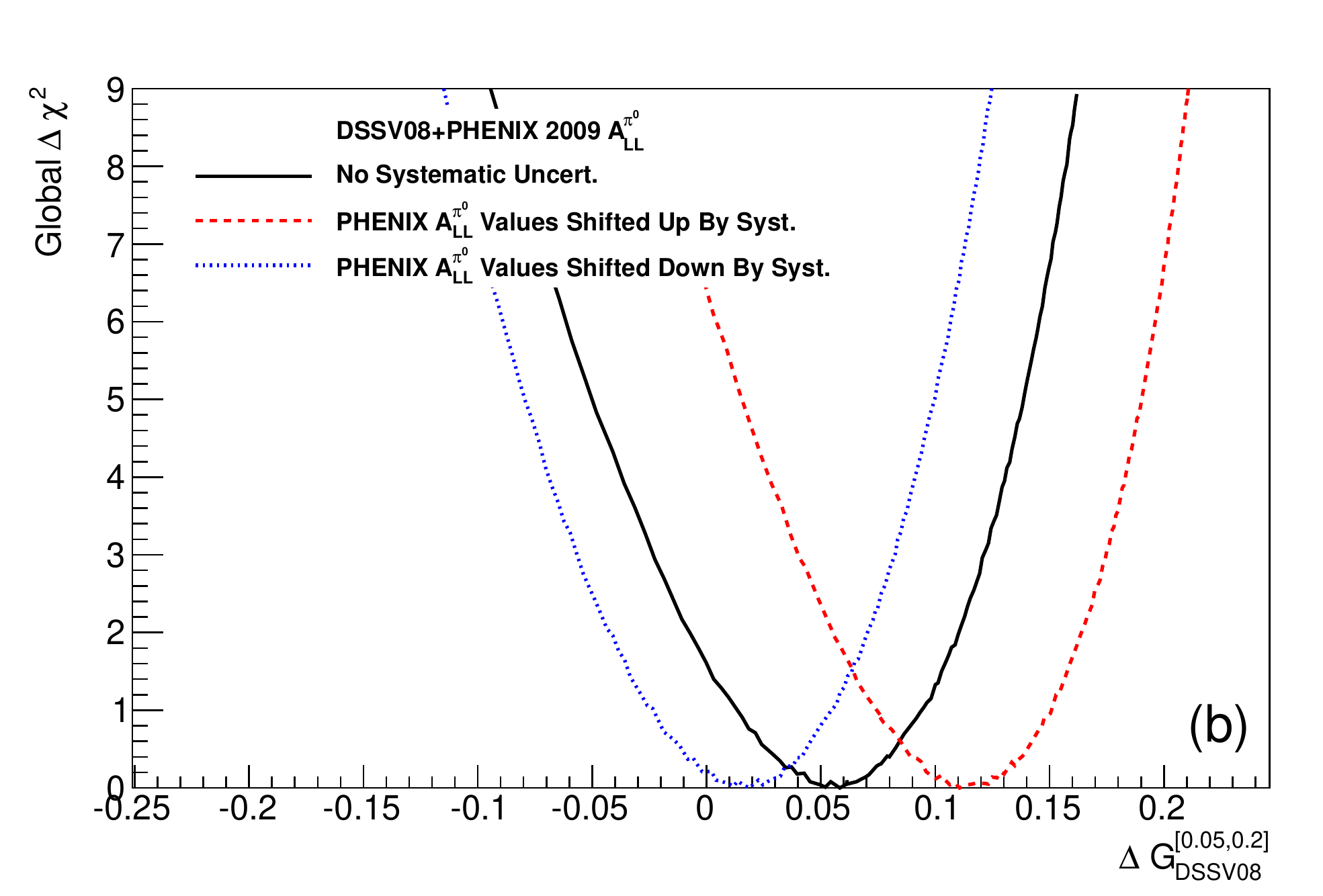}
\caption{(color online)
Change in $\chi^2$ as a function of 
$\Delta G^{[0.05,0.2]}_{\mbox{\sc dssv08}}$, when adding PHENIX 2009 \pz data 
to the {\sc dssv08} global analysis (which includes earlier PHENIX \pz data).  
(a) Contribution of the combined PHENIX data at $\sqs = 200$~GeV to the 
global $\chi^2$ of the {\sc dssv08} analysis using only statistical 
uncertainties.  The different curves show the effect of shifting only the 
PHENIX data points up or down by their total systematic uncertainty, which 
is $p_T$ correlated.  (b) The effect of shifting only the PHENIX $\sqs = 
200$~GeV data points on the {\sc dssv08} \emph{global} $\chi^2$.}
\label{f:DG_plusminus_syst}
\end{figure*}

In Figs.~\ref{f:pi0alltheocomp} and \ref{f:etaalltheocomp}, the 2005, 2006 
and 2009 results have been combined for the \pz and \et, respectively, 
with the uncorrelated part of the systematic uncertainties combined and 
shown on the points.  The year-to-year correlated parts of the 
polarization and relative luminosity uncertainties are given in the 
legend.

Both the \et and \pz asymmetries are consistent with the best fit of a 
global analysis of DIS data that allows at the input scale only quark 
contributions to \all: the GRSV-zero scenario, which assumes $\Delta 
g(x,\mu^2) = 0$ at an NLO input scale $\mu^2 = 0.40 \mbox{~GeV}^2$ 
\cite{ref:GRSV}.  This consistency can be quantified relative to the 
related GRSV-std scenario, in which the gluon polarization is not fixed 
(nor is it well constrained).  The difference between these two scenarios 
in a statistical-uncertainty-only comparison to the combined \pz data in 
the 2--9~\punit \pt range is $ (\Delta \chi^2)_{{\rm GRSV}} / 
\mbox{N.D.F} \equiv \left( (\chi^2)_{{\rm GRSV-std}} - 
(\chi^2)_{{\rm GRSV-zero}} \right) / \mbox{N.D.F} = 18.9/8$, a 
$4.3$-sigma change.  If all of the points are increased by the total 
systematic uncertainty to move them closer to the GRSV-std curve, the 
change is $3.3/8$ or $1.8$ sigma, indicating that the PHENIX \pz data 
still prefer the GRSV-zero scenario.  For the \et asymmetries in the same 
\pt range, $(\Delta \chi^2)_{{\rm GRSV}} / \mbox{N.D.F} = 0.4/6$ or $0.6$ 
sigma, indicating a slight preference for GRSV-zero.  With the \et 
asymmetries shifted up by the systematic uncertainty, there is a slight 
preference for GRSV-std, with $(\Delta \chi^2)_{{\rm GRSV}} / 
\mbox{N.D.F} = -0.1/6$ or $0.3$ sigma.

More recent NLO global analyses of DIS-only data by Bl{\"u}mlein and 
B{\"o}ttcher (BB10)~\cite{ref:BB10} and Ball et. al. 
(NNPDF)~\cite{ref:NNPDF_pol, ref:NNPDF_unpol}, and of DIS+SIDIS data by 
Leader et. al. (LSS10)~\cite{ref:LSS} also allow the gluon polarization to 
be fit by the data, but the analyses vary in ways that affect 
determination of $\Delta g(x,\mu^2)$.  The most significant of these 
differences is the BB10 assumption of a flavor-symmetric sea versus the 
separation of flavor-specific distributions made possible in LSS10 by the 
SIDIS data.  This affects the gluon determination not only because of the 
constraint on the total polarization, but also because the analyses use 
functional forms for the initial pPDFs such as
\begin{equation}
  x \Delta f_i (x,\mu^2) = N_i x^{\alpha_i}(1-x)^{\beta_i}(1+\gamma_i \sqrt{x} + \eta_i x)
\label{eq:pdf_form}
\end{equation}
and consequentially must relate parameters between the sea and gluon 
distributions to enforce positivity ($|\Delta f_i (x,\mu^2)| \leq f_i 
(x,\mu^2)$) and to fix poorly-constrained parameters.

Another issue with making a choice of functional form for $\Delta 
g(x,\mu^2)$ is that, even with inclusion of present $\vec{p}+\vec{p}$ 
data, there are no existing measurements that can test the validity of the 
functional form in the low-$x$ region.  For analyses like BB10 and LSS10 
that do not include $\vec{p}+\vec{p}$ data, this problem extends to 
determination of $\Delta g$ in the medium and large-$x$ regions as well.  
The NNPDF analysis of DIS data avoids bias introduced in choosing a 
functional form for the PDFs by using neural networks to control 
interpolation between different $x$ values.  For example, $\Delta 
g(x,\mu^2)$ is parameterized as
\begin{equation}
  \Delta g(x,\mu^2) = (1-x)^m x^{-n} NN_{\Delta g}(x),
\end{equation}
with $NN_{\Delta g}(x)$ a neural network parameterization determined by 
scanning functional space for agreement with 1000 randomly distributed 
replicas of the experimental data.  The low- and high-$x$ terms are 
included for efficiency, and to ensure that they do not bias the 
fit, $m$ and $n$ are chosen from a random interval for each experimental 
data replica such that this interval is wider than the range of effective 
exponents for the limiting low and high-$x$ behavior after the neural 
network terms have been included.

Fig.~\ref{f:pi0alltheocomp}(b) includes $A_{LL}^{\pz}$ predictions based 
on the BB10, LSS10, and NNPDF polarized PDF determinations.  For BB10 and 
LSS10, we evolved their published polarized PDFs to various $\mu^2$ using 
the QCD-PEGASUS package~\cite{ref:PEGASUS} and used these to calculate the 
$p_T$ dependent polarized cross section for inclusive \pz production with 
code based on \cite{ref:xsect_code} that uses the DSS NLO fragmentation 
functions~\cite{ref:DSS_FF}.  The unpolarized cross section for the 
denominator was calculated via the same two-step process starting from the 
CTEQ-6 PDFs~\cite{ref:CTEQ_6}.  The BB10 uncertainty band was calculated 
using the Hessian method with a set of polarized PDFs obtained from the 
parameter covariance matrix in the BB10 publication.  The NNPDF prediction 
was provided by that group, using their polarized PDFs supplemented by 
preliminary W boson asymmetry measurements from the STAR 
experiment~\cite{ref:NNPDF_W, ref:STAR_W}.  Neither the BB10 nor NNPDF 
prediction accounts for uncertainties in the determination of the \pz 
fragmentation functions.

One feature of the predictions is that the BB10 uncertainty band is 
smaller than the NNPDF band at $\pt \approx 3$~\punit but quickly exceeds 
it as \pt increases.  Likewise, as can be seen in 
Ref.~\cite{ref:NNPDF_pol}, at an input scale of 4~\gunit, the 
uncertainty on the BB10 prediction for $\Delta g$, which neglects bias 
from the choice of functional form, is smaller than that for NNPDF at 
low-$x$ but exceeds it as $x$ increases.  Future inclusion of the PHENIX 
$A_{LL}^{\pz}$ into the NNPDF analysis may provide some insight into 
whether or not this is due to a bias in the choice of functional form at 
medium-$x$, particularly in the RHIC range of $[0.05,0.2]$.

The {\sc dssv08} global analysis \cite{ref:DSSV2}, which is also based on the 
pPDF parameterizations of Eq.~\ref{eq:pdf_form}, includes, in addition to 
DIS and SIDIS data, final 2005 RHIC data 
\cite{ref:run5pi0xsect_all,ref:run5starjetall} and preliminary versions of 
the 2006 RHIC data presented in \cite{ref:run6starjetxsect_all, 
ref:run6pi0_all, ref:vern}.  The results of that analysis, which yields a 
much more accurate determination of $\Delta g(x)$, are compared with 
$A_{LL}^{\pz}$ in Fig.~\ref{f:pi0alltheocomp}(a).  We also ran an updated 
version of the {\sc dssv08} analysis to include final versions of the RHIC data 
through 2006 \cite{ref:run6starjetxsect_all, ref:run6pi0_all, ref:vern} 
along with the final $A_{LL}^{\pz}$ results presented here.  We obtained 
$\Delta G^{[0.05,0.2]}_{\mbox{\sc dssv08}} = 0.06^{+0.04}_{-0.06}(\Delta 
\chi^2 = 1)^{+0.11}_{-0.15}(\Delta \chi^2 = 9)$, where the $\Delta \chi^2 
= 9$ uncertainties roughly correspond to the $2\%$ change in $\Delta 
\chi^2/\chi^2_{\rm min}$ used to determine the uncertainties in the {\sc dssv08} 
global analysis.  The full $\Delta \chi^2$ curve from our updated analysis 
is shown as the central curve in Fig.~\ref{f:DG_plusminus_syst}(b).  
Fig.~\ref{f:DG_plusminus_syst}(a) shows the contribution from PHENIX data 
to that curve, and that data prefers $\Delta 
G^{[0.05,0.2]}_{\mbox{PHENIX}} = 0.07^{+0.05}_{-0.08}(\Delta \chi^2 = 1)$.

Systematic uncertainties for the RHIC dataset were not included in the 
{\sc dssv08} analysis.  However, the PHENIX relative-luminosity systematic 
uncertainty now exceeds the statistical uncertainty on $A_{LL}^{\pz}$ in 
the lowest $p_T$ bins.  To understand the impact of this on the fit 
result, we shifted the PHENIX $\sqs = 200$~GeV data up and down by the 
systematic uncertainties given in the final column of 
Table~\ref{t:pi0_combined_datatable}, while ignoring the systematic 
uncertainties of all other data sets.  As demonstrated in 
Fig.~\ref{f:DG_plusminus_syst}, this changes the global best-fit value to 
$0.12$ or $0.02$, with the value preferred by the PHENIX data changing to 
$0.17$ or $-0.03$.  It is therefore necessary to include this uncertainty 
in future global analyses to obtain accurate determinations of 
$\Delta G$.

\section{Summary}

We present the latest PHENIX measurements of \all in \pz and \et 
production in longitudinally polarized $p$$+$$p$ collisions at $\sqs=200$~GeV. 
These results are compared with various existing DIS and SIDIS global 
analyses \cite{ref:GRSV, ref:BB10, ref:LSS, ref:NNPDF_pol, 
ref:NNPDF_unpol} and found to be consistent within the fit uncertainties.  
We also find consistency with the {\sc dssv08} global analysis \cite{ref:DSSV1, 
ref:DSSV2}, which includes versions of earlier PHENIX measurements.  
Addition of our new results to that analysis (as well as the updating of 
previous RHIC data~\cite{ref:run6starjetxsect_all, 
ref:run6pi0_all, ref:vern}) yields a statistical-uncertainty only 
constraint of 
$\Delta G^{[0.05,0.2]}_{\mbox{\sc dssv08}} = 0.06^{+0.11}_{-0.15}$ 
with uncertainties determined at $\Delta \chi^2 = 9$.  However, we 
emphasize the importance of including the relative-luminosity systematic 
uncertainty in future analyses that use RHIC asymmetries, since shifting 
the $\sqrt{s} = 200$~GeV PHENIX data alone down and up by its systematic 
uncertainty changes the global best-fit value 
$\Delta G^{[0.05,0.2]}_{\mbox{\sc dssv08}}$ from $0.02$ to $0.12$.  A 
significant effort by the RHIC experiments to understand and correct for 
the relative-luminosity systematic effect is also currently underway.  
Furthermore, for the \et asymmetries to be used, better determination of 
\et fragmentation functions is needed, perhaps using the well-determined 
\pz to \et cross-section ratio \cite{ref:run6etaxsect_all, 
ref:eta_pi0_xsect_ratios}.


\section*{ACKNOWLEDGMENTS}


We thank the staff of the Collider-Accelerator and Physics
Departments at Brookhaven National Laboratory and the staff of
the other PHENIX participating institutions for their vital
contributions.  We acknowledge support from the
Office of Nuclear Physics in the
Office of Science of the Department of Energy, the
National Science Foundation, Abilene Christian University
Research Council, Research Foundation of SUNY, and Dean of the
College of Arts and Sciences, Vanderbilt University (U.S.A),
Ministry of Education, Culture, Sports, Science, and Technology
and the Japan Society for the Promotion of Science (Japan),
Conselho Nacional de Desenvolvimento Cient\'{\i}fico e
Tecnol{\'o}gico and Funda\c c{\~a}o de Amparo {\`a} Pesquisa do
Estado de S{\~a}o Paulo (Brazil),
Natural Science Foundation of China (P.~R.~China),
Ministry of Education, Youth and Sports (Czech Republic),
Centre National de la Recherche Scientifique, Commissariat
{\`a} l'{\'E}nergie Atomique, and Institut National de Physique
Nucl{\'e}aire et de Physique des Particules (France),
Bundesministerium f\"ur Bildung und Forschung, Deutscher
Akademischer Austausch Dienst, and Alexander von Humboldt Stiftung (Germany),
Hungarian National Science Fund, OTKA (Hungary),
Department of Atomic Energy and Department of Science and Technology (India),
Israel Science Foundation (Israel),
National Research Foundation and WCU program of the
Ministry Education Science and Technology (Korea),
Physics Department, Lahore University of Management Sciences (Pakistan),
Ministry of Education and Science, Russian Academy of Sciences,
Federal Agency of Atomic Energy (Russia),
VR and Wallenberg Foundation (Sweden),
the U.S. Civilian Research and Development Foundation for the
Independent States of the Former Soviet Union,
the Hungarian American Enterprise Scholarship Fund,
and the US-Israel Binational Science Foundation.




%

\end{document}